\documentclass[11pt]{article}
\pdfoutput=1

\usepackage{jheppub}
\usepackage[T1]{fontenc}

\usepackage{amssymb}
\usepackage{amsfonts}
\usepackage{amsbsy}
\usepackage{amsmath}
\usepackage{graphicx}
\usepackage{epstopdf}
\def\Z{\mathbb{Z}}
\def\F{\mathbb{F}}
\def\Q{\mathbb{Q}}
\def\C{\mathbb{C}}

\def\P{\mathbb{P}}
\def\E{\mathbb{E}}

\def\n3a{t}

\def\tr{{\mathrm{tr}}}

\def\ge{{\mathfrak{e}}}
\def\gso{{\mathfrak{so}}}
\def\gsu{{\mathfrak{su}}}
\def\gsp{{\mathfrak{sp}}}
\def\gf{{\mathfrak{f}}}
\def\gg{{\mathfrak{g}}}

\newcommand{\eq}[1]{(\ref{#1})}

\title{Calabi-Yau threefolds with large $h^{2, 1}$}

\author{Samuel B.\ Johnson and Washington Taylor}

\affiliation{Center for Theoretical Physics\\
Department of Physics\\
Massachusetts Institute of Technology\\
77 Massachusetts Avenue\\
Cambridge, MA 02139, USA}

\emailAdd{{\tt samj} {\rm at} {\tt mit.edu},
{\tt wati} {\rm at} {\tt mit.edu}}

\preprint{MIT-CTP-4535}

\abstract{
We carry out a systematic analysis of Calabi-Yau threefolds that are
elliptically fibered with section (``EFS'') and have a large Hodge
number $h^{2, 1}$.  EFS Calabi-Yau threefolds live in a single
connected space, with regions of moduli space associated with
different topologies connected through transitions that can be
understood in terms of singular Weierstrass models.  We determine the
complete set of such threefolds that have $h^{2, 1} \geq 350$ by
tuning coefficients in Weierstrass models over Hirzebruch surfaces.
The resulting set of Hodge numbers includes those of all known
Calabi-Yau threefolds with $h^{2, 1} \geq 350$, as well as three
apparently new Calabi-Yau threefolds.  We speculate that there are no
other Calabi-Yau threefolds (elliptically fibered or not) with Hodge
numbers that exceed this bound.  We summarize the theoretical and
practical obstacles to a complete enumeration of all possible EFS
Calabi-Yau threefolds and fourfolds, including those with small Hodge
numbers, using this approach.}

\begin{document}
\maketitle

\flushbottom

%--------------------------------
\section{Introduction}
\label{sec:intro}

Since the early days of string theory, the geometry of Calabi-Yau (CY)
threefolds has played an important role in understanding
compactifications of the theory that give rise to four-dimensional
effective physics \cite{chsw, GSW, Polchinski}.  Over the last three
decades, progress from both mathematical and physical directions has
led to the construction of a wide range of specific Calabi-Yau
threefold geometries, and some general results on the structure of
these manifolds.  (See for example \cite{Hubsch, ghj, Davies, He-CY}.)  Many basic
questions regarding this class of geometries remain unanswered,
however, such as whether the number of distinct topological types of
CY threefolds is finite.

The class of CY threefolds that admit an elliptic ($T^2$ with complex
structure) fibration with at least one section forms a subset of the
full set of Calabi-Yau manifolds that is of interest both for
mathematical and physical reasons.  Mathematically, the existence of
an elliptic fibration adds structure that simplifies the analysis and
classification of possible CY geometries. It has been proven by Gross
\cite{Gross} that there are a finite number of distinct topological
types (up to birational equivalence) of elliptically fibered
Calabi-Yau threefolds.  For an elliptically fibered Calabi-Yau
threefold, the existence of a global section makes possible an
explicit presentation as a Weierstrass model \cite{Nakayama}
\begin{equation}
 y^2 = x^3+ f x+ g \,, 
\label{eq:Weierstrass-1}
\end{equation}
where $f, g$ are functions (really, sections of line bundles) on the
base $B_2$ of the elliptic fibration $\pi: X_3 \rightarrow B_2$,
$\pi^{-1} (p) \cong \E \cong T^2$.  Elliptically-fibered CY threefolds
with section (henceforth ``EFS CY3s'') have a role in physics as
compactification spaces for F-theory \cite{Vafa-F-theory,
  Morrison-Vafa-I} 
that give rise to six-dimensional theories of
supergravity.  F-theory can be thought of as a nonperturbative
description of type IIB string theory where the axiodilaton field
$\chi +i e^{- \phi}$ varies over a compact space (a complex surface
$B_2$ for supersymmetric 6D theories) and parameterizes the elliptic
curve over this base.  One particularly nice feature of the set of
elliptically fibered Calabi-Yau threefolds is that their moduli spaces
are all connected through singular Weierstrass models.  On the physics
side this unifies all the corresponding F-theory vacuum solutions of 6D
supergravity into a single theory;
tensionless string transitions \cite{Seiberg-Witten,
  Morrison-Vafa-II} connect the different branches of the theory.  In addition to the special features that make
them easier to analyze mathematically and connect them to the physics
of F-theory, elliptically fibered threefolds may comprise a large
fraction of the set of {\em all} Calabi-Yau threefolds, particularly
those with large Hodge numbers.  This paper contributes to a growing
body of
circumstantial evidence for this conclusion,
which is discussed further in \S\ref{sec:role-EFS}.

The close connection between the physics of 6D supergravity theories
and the geometry of EFS Calabi-Yau threefolds leads to a physically
motivated approach to the classification of these geometries. From the
work of Grassi \cite{Grassi} and the mathematical minimal model
program for classifying surfaces \cite{bhpv, Reid-chapters}, it is
known that all complex surfaces $B_2$ that support elliptically
fibered Calabi-Yau threefolds are in the set consisting of $\P^2,$ the
Enriques surface, the Hirzebruch surfaces $\F_m$ for $0 \leq m \leq
12$, and blow-ups of the $\F_m$ at one or more points.  As argued in
\cite{KMT-II}, the finiteness of the set of EFS CY3s can then be
understood in a constructive context from the finite number of
topologically distinct tunings (strata) of the class of Weierstrass
models over the minimal bases $\P^2$ and $\F_m$ (the Enriques surface
is not as interesting since, up to torsion, the canonical class $K$
vanishes, so the Weierstrass model is essentially trivial).  From this
point of view, in principle all EFS CY threefolds can be constructed
by starting with the base surfaces $\P^2$ and $\F_m$, and tuning the
Weierstrass models over these bases in all possible ways consistent
with the existence of a Calabi-Yau elliptic fibration.  The set of
such possible tunings can be described conveniently in terms of the
spectra (gauge group and matter content) of the corresponding 6D
supergravity theories.  A complete classification of the types of
intersection structures (corresponding to ``non-Higgsable clusters''
in the physical picture) that can appear in the base $B_2$ was given
in \cite{clusters}.  This was used in \cite{toric} to explicitly
construct all toric bases $B_2$ that support EFS CY3s, and in
\cite{Martini-WT} to construct a broader class of bases admitting a
single $\C^*$ action.  The generic elliptic fibrations over these
toric and ``semi-toric'' bases were shown \cite{WT-Hodge} to
include the EFS CY3 with the largest possible value of the Hodge
number $h^{2, 1}$ (= 491), and to describe in outline the
``shield-shaped'' boundary on the set of known Hodge numbers found
experimentally by Kreuzer and Skarke \cite{Kreuzer-Skarke}.

In this paper we pursue this line of inquiry further by explicitly
constructing all EFS CY3s with large $h^{2, 1}$ through the tuning of
Weierstrass models on $\F_m$ for $m \geq 7$.  We systematically
construct all CY3s with $h^{2, 1} \geq 350$, and compare with known
data for threefolds with large $h^{2, 1}$.  The Hodge number pairs for
EFS Calabi-Yau threefolds that have $h^{2, 1}\geq 350$ are plotted in
Figure~\ref{f:results}; the detailed structure and construction of
these threefolds is explained in the bulk of the paper.  The arbitrary
bound of 350 is chosen so that the number of possible threefolds is
both limited enough to be manageable in a case-by-case analysis, and
rich enough to illustrate the range of principles involved.  A similar
analysis can be used to systematically construct all elliptically
fibered Calabi-Yau threefolds with section at increasingly small
values of $h^{2, 1}$.  While this procedure becomes computationally
intensive at lower values of $h^{2, 1}$, and there are a number of
practical and theoretical issues that must be resolved before a
complete classification is possible, this program could in principle
be pursued to enumerate {\it all} EFS CY3s.

The structure of this paper is as follows: In \S\ref{sec:tools} we
review the basic structure of EFS CY3s and describe some aspects of
the Weierstrass tunings needed to construct explicit CY3s.  In
\S\ref{sec:systematic} we give a complete classification of all EFS CY3s
that have $h^{2, 1} \geq 350$.  We conclude in \S\ref{sec:conclusions}
with a description of the technical obstructions to classifying all
EFS Calabi-Yau threefolds, and give a summary of results and
discussion of further directions.

\section{Classification of CY threefolds that are
  elliptically fibered with section}
\label{sec:tools}

\subsection{Elliptic fibrations and F-theory}

As summarized in \S\ref{sec:intro}, the Weierstrass form of an
elliptic fibration $y^2=x^3+fx z^4+g z^6$ describes the total space of
a Calabi-Yau threefold $X$ in terms of information on the base surface
(complex twofold) $B_2$ by determining the complex structure of the elliptic
fiber (torus) $\E \cong T^2$ over each point in the base in terms of a
(complex) curve in $\P^{2,3, 1}$. More precisely,  $f$
and $g$, as well as the discriminant $\Delta = 4f^3+27g^2$, are
sections of line bundles 
\[
f \in{\cal O} (-4K)\ \ \ g \in{\cal O} (-6K)\ \ \ \Delta \in 
{\cal O} (-12K) \,,
\]
where $K$ is the canonical
class of the base $B_2$.  Note that throughout this
paper we will be somewhat informal about the distinction between
divisors $D$ in $B_2$ and the associated homology classes $[D]$ in $H_2
(B_2,\Z)$.

The singularity structure of $X$ as an elliptic fibration over $B_2$
is encoded in the vanishing loci of $f, g,$ and $\Delta$.  The close
correspondence between the geometry of an elliptic fibration and the
corresponding physical F-theory model illuminates both the
mathematical and physical properties of these constructions.
(Pedagogical introductions to F-theory compactifications can be found
in \cite{Morrison-TASI, Denef-F-theory, WT-TASI}.)  The
codimension one loci where $\Delta$ vanishes to higher degree lead to
singularities in the total space of the threefold that must be
resolved to form a smooth Calabi-Yau total space.  These singularities
correspond physically to 7-branes in the F-theory picture, and the
degrees of vanishing of $f, g, \Delta$ (along with monodromy
information in some cases) encode geometric structure that corresponds
to the Lie algebra of the nonabelian gauge group $G$ of the 6D theory according to the Kodaira-Tate
classification of singularities summarized in Table~\ref{t:Kodaira}
\cite{Kodaira, Morrison-Vafa-II, Bershadsky-all, Morrison-sn,
  Grassi-Morrison-2}.  The codimension two vanishing loci of $\Delta$
encode further singularities associated with hypermultiplet matter in
the 6D theory transforming under some combination of irreps of the
gauge group $G$.  While the correspondence between matter and
codimension two singularities is understood in the simplest and most
generic cases, there is not yet a complete dictionary of this
correspondence for general matter representations and arbitrary
codimension two singularities.  A further discussion of exotic matter
representations and associated singularities appears in
\S\ref{sec:issues-matter}.

\begin{table}
\begin{center}
\begin{tabular}{|c |c |c |c |c |c |}
\hline
Type &
ord ($f$) &
ord ($g$) &
ord ($\Delta$) &
singularity & nonabelian symmetry algebra\\ \hline \hline
$I_0$&$\geq $ 0 & $\geq $ 0 & 0 & none & none \\
$I_n$ &0 & 0 & $n \geq 2$ & $A_{n-1}$ & $\gsu(n)$  or $\gsp(\lfloor
n/2\rfloor)$\\ 
$II$ & $\geq 1$ & 1 & 2 & none & none \\
$III$ &1 & $\geq 2$ &3 & $A_1$ & $\gsu(2)$ \\
$IV$ & $\geq 2$ & 2 & 4 & $A_2$ & $\gsu(3)$  or $\gsu(2)$\\
$I_0^*$&
$\geq 2$ & $\geq 3$ & $6$ &$D_{4}$ & $\gso(8)$ or $\gso(7)$ or $\gg_2$ \\
$I_n^*$&
2 & 3 & $n \geq 7$ & $D_{n -2}$ & $\gso(2n-4)$  or $\gso(2n -5)$ \\
$IV^*$& $\geq 3$ & 4 & 8 & $\ge_6$ & $\ge_6$  or $\gf_4$\\
$III^*$&3 & $\geq 5$ & 9 & $\ge_7$ & $\ge_7$ \\
$II^*$& $\geq 4$ & 5 & 10 & $\ge_8$ & $\ge_8$ \\
\hline
non-min &$\geq 4$ & $\geq6$ & $\geq12$ & \multicolumn{2}{c|}{ does not occur in 
F-theory } \\ 
\hline
\end{tabular}
\end{center}
\caption[x]{\footnotesize  Table of 
codimension one
singularity types for elliptic
fibrations and associated nonabelian symmetry algebras.
In cases where the algebra is not determined uniquely by the degrees
of vanishing of $f, g$,
the precise gauge algebra is fixed by monodromy conditions that can be
identified from the form of the Weierstrass model.
}
\label{t:Kodaira}
\end{table}

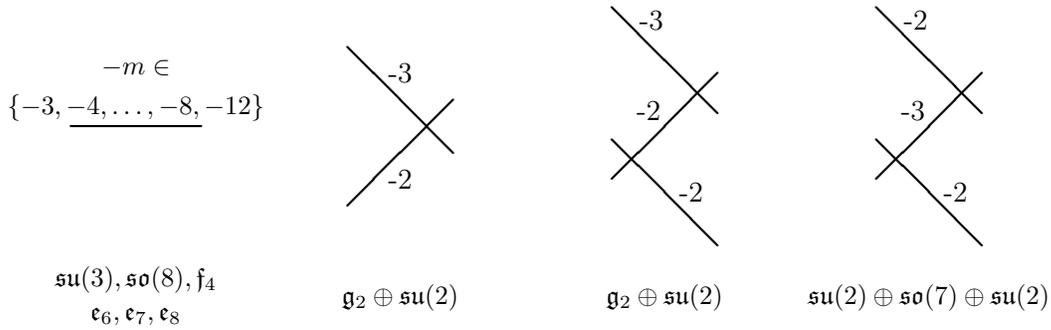
\begin{figure}
\begin{center}
\begin{picture}(200,130)(- 93,- 55)
%\grid
\thicklines
\put(-175, 25){\line(1,0){50}}
%\put(-150,32){\makebox(0,0){$-m\leq -3$}}
\put(-150, 47){\makebox(0,0){\small $-m \in$}}
\put(-150, 32){\makebox(0,0){\small  $\{-3, -4, \ldots, -8, -12\}$}}
%\put(-150, 18){\makebox(0,0){$(m = 3, 4, 5, 6, 7, 8, 12)$}}
\put(-150,-33){\makebox(0,0){\small $\gsu(3), \gso(8), \gf_4$}}
\put(-150,-47){\makebox(0,0){\small $\ge_6, \ge_7, \ge_8$}}
\put(-70,55){\line(1,-1){40}}
\put(-30,35){\line(-1,-1){40}}
\put(-50,45){\makebox(0,0){-3}}
\put(-50,5){\makebox(0,0){-2}}
\put(-50,-40){\makebox(0,0){\small $\gg_2 \oplus \gsu(2)$}}
\put(30,70){\line(1,-1){40}}
\put(30,20){\line(1,-1){40}}
\put(70,45){\line(-1,-1){40}}
\put(45,65){\makebox(0,0){-3}}
\put(44,31){\makebox(0,0){-2}}
\put(60, 0){\makebox(0,0){-2}}
\put(50,-40){\makebox(0,0){\small $\gg_2 \oplus \gsu(2)$}}
\put(130,70){\line(1,-1){40}}
\put(130,20){\line(1,-1){40}}
\put(170,45){\line(-1,-1){40}}
\put(145,65){\makebox(0,0){-2}}
\put(144,31){\makebox(0,0){-3}}
\put(160, 0){\makebox(0,0){-2}}
\put(150,-40){\makebox(0,0){\small $\gsu(2) \oplus \gso(7) \oplus \gsu(2)$}}
\end{picture}
\end{center}
\caption[x]{\footnotesize    Clusters of intersecting
  curves that must carry a nonabelian gauge group factor.  For each
cluster the corresponding gauge algebra is noted and the gauge algebra and
number of charged matter hypermultiplet are listed in Table~\ref{t:clusters}}
\label{f:clusters}
\end{figure}

The generic Weierstrass model over a given base $B_2$ may have
singularities in the elliptic fibration that are forced by the
structure of irreducible effective divisors (curves) of negative
self-intersection in $B_2$ over which $f, g,$ and $\Delta$ must
vanish.  The possible configurations of curves that give rise to
mandatory singularities -- corresponding to ``non-Higgsable clusters''
of gauge groups and possible matter in the 6D F-theory picture -- were
classified in \cite{clusters} and are depicted in
Figure~\ref{f:clusters}, with the minimal gauge group and matter
content for each cluster listed in Table~\ref{t:clusters}.  By further
tuning the coefficients in the Weierstrass model, higher degree
singularities can be produced on $f, g,$ and $\Delta$, corresponding
to enhanced gauge groups in the 6D supergravity theory.  In this way,
a variety of topologically distinct Calabi-Yau threefolds can be
constructed by tuning the Weierstrass model over a given base $B_2$.

\begin{table}
\begin{center}
\begin{tabular}{| c | 
c | c |c |c |
}
\hline
Cluster & gauge algebra & $r$ & $V$ &  $H_{\rm charged}$ 
\\
\hline
(-12) &$\ge_8$ & 8 & 248 &0 \\
(-8) &$\ge_7$&  7 &  133 &0 \\
(-7) &$\ge_7$& 7 & 133 &28 \\
(-6) &$\ge_6$&   6 & 78 &0 \\
(-5) &$\gf_4$&   4 & 52 &0 \\
(-4) &$\gso(8) $&  4 & 28 &0 \\
(-3, -2, -2)  &  $\gg_2 \oplus \gsu(2)$&  3 & 17 &8\\
(-3, -2) &  $\gg_2 \oplus \gsu(2)$ &3 & 17 &8 \\
(-3)& $\gsu(3)$ &  2 & 8 &0 \\
(-2, -3, -2) &$\gsu(2) \oplus \gso(7) \oplus
\gsu(2)$& 5 &  27 &16 \\
(-2, -2, \ldots, -2) & no gauge group & 0 & 0 &0 \\
\hline
\end{tabular}
\end{center}
\caption[x]{\footnotesize
List of ``non-Higgsable clusters'' of
  irreducible effective divisors with self-intersection $-2$ or below,
  and corresponding contributions to the gauge algebra and matter
  content of the 6D theory associated with F-theory compactifications
  on a generic elliptic fibration (with section) over a base
  containing each cluster.
The quantities $r$ and $V$ denote the rank and dimension of the
nonabelian gauge algebra, and $H_{\rm charged}$ denotes the number of
charged hypermultiplet matter fields associated with intersections
between the curves supporting the gauge group factors.
}
\label{t:clusters}
\end{table}

In general, a smooth Calabi-Yau threefold can be constructed by
resolving all singularities in the total space of the elliptic
fibration.  If the vanishing of $f, g, \Delta$ reaches degrees (4, 6,
12) over a divisor then there is no smooth resolution of the singular
elliptic fibration as a Calabi-Yau threefold.  If the vanishing of $f,
g, \Delta$ reaches degrees (4, 6, 12) on a codimension two locus in
the base (a point in the surface $B_2$), then the point must be blown
up to form a new base ${B}'_2$ over which there is a smooth CY3
after resolution of singularities (unless the blow-up leads to
additional (4, 6, 12) vanishing on a divisor or point in the new
base). Thus, to describe the possible EFS CY3s over a given base
$B_2$, we need only consider tunings where the degrees of vanishing of
$f, g, \Delta$ do not reach (4, 6, 12) at any point in the base.

For any given EFS CY threefold $X$ with a Weierstrass description over
a given base $B_2$, the Hodge numbers of $X$ can be read off from the
form of the singularities.  A succinct description  of the Hodge
numbers of $X$ can be given using the geometry-F-theory correspondence
\cite{Morrison-Vafa-I, Morrison-Vafa-II, WT-Hodge} 
\begin{eqnarray}
h^{1, 1}(X) & = & r + T +2 \label{eq:11}\\
h^{2, 1} (X) & = &  H_{\rm  neutral}-1 = 272+ V -29T - H_{\rm  charged}
\,.
\label{eq:21}
\end{eqnarray}
Here, $T = h^{1, 1} (B_2) -1$ is the number of tensor multiplets in
the 6D theory; $r$ is the rank of the 6D gauge group and $V$ is the
number of vector multiplets in the 6D theory, while $H_{\rm neutral}$
and $H_{\rm charged}$ refer to the number of 6D matter hypermultiplets
that are neutral/charged with respect to the gauge group $G$.  The
relation (\ref{eq:11}) is essentially the Shioda-Tate-Wazir formula
\cite{stw}.  The equality (\ref{eq:21}) follows from the gravitational
anomaly cancellation condition in 6D supergravity, $H - V = 273-29T,$ which
corresponds to a topological relation on the Calabi-Yau side that has
been verified for most matter representations with known nongeometric
counterparts \cite{Grassi-Morrison, Grassi-Morrison-2}. The
nonabelian part of the gauge group $G$ can be read off from the
Kodaira types of the singularities in the elliptic fibration according
to Table~\ref{t:Kodaira}
(up to a discrete part that does not affect the Hodge numbers and that
we do not compute in detail here.).  The contribution to the rank $r$ and the
numbers of vector multiplets $V$ and charged hypermultiplets $H_{\rm
  neutral}$ for the gauge fields and matter associated with
non-Higgsable clusters (NHCs) are listed in Table~\ref{t:clusters}.

In principle, $G$ can also have abelian (U(1)) factors, corresponding
to additional sections of the elliptic fibration, which contribute to $h^{1, 1}(X)$ through $r$.  Mathematically,
these sections are associated with a higher rank Mordell-Weil group of
the fibration.  There can also be torsion in the Mordell-Weil group
\cite{mx-torsion},
which corresponds to the discrete part of the gauge group in the gravity theory, but
does not contribute to the Hodge numbers of the elliptic fibration.
Because abelian factors arise from global, rather than local, aspects
of the total space of the elliptic fibration, it is difficult to
systematically describe U(1) factors in the gauge group.  Though there
has been substantial progress on this problem in recent years, as we
show in \S\ref{sec:tuning-0}, abelian $U(1)$
factors
in the gauge group cannot arise for EFS CY3s with $h^{2, 1} \geq 350$,
so we do not need to consider them in this paper, and $r$ and $V$ in
(\ref{eq:11}, \ref{eq:21}) can be read off directly from the
codimension one singularities in the elliptic fibration. The
representations and multiplicity of charged matter needed to compute
$H_{\rm charged}$ in (\ref{eq:21}) can also be read off directly from
the form of the local singularities in the absence of abelian gauge
group factors.  While the types of singularities associated with
completely general matter representations have not yet been
classified, the codimension two singularities that arise in EFS CY3s
of large $h^{2, 1}$ belong to the simple categories of
well-understood matter representations and associated singularities.

We now describe some of the details of the steps needed to
systematically classify EFS Calabi-Yau threefolds starting at large
$h^{2, 1}$.

\subsection{Systematic classification of EFS Calabi-Yau threefolds}

A complete classification and enumeration of Calabi-Yau threefolds that are
elliptically fibered with section can in principle be carried out in
three steps:

\begin{enumerate}
\item
Classify and enumerate all bases $B_2$ that support a smooth
elliptically fibered 
Calabi-Yau threefold with section.
\item
Classify and enumerate all codimension one gauge groups that can be
``tuned'' over a given base, giving enhanced gauge groups in the 6D
theory.
\item
Given the gauge group structure, classify and enumerate the set of
compatible matter representations -- in some cases this may involve
further tuning of codimension two singularities.
\end{enumerate}

In the remainder of this section we describe some general aspects of
the procedures involved in these steps 1--3 for the construction of
EFS CY3s with large $h^{2, 1}$.
Some of the technical limitations to carrying out these three steps
for all EFS Calabi-Yau threefolds are discussed in \S\ref{sec:limitations-3}.

A key principle that enables efficient classification of the
threefolds of interest through the structure of their singularities is
the decomposition of an effective divisor (curve) $D$ in $B_2$ into a
{\it  base locus} of irreducible effective curves $C_i$ of negative
self-intersection, and a residual part $X$, which satisfies $X \cdot C
\geq 0$ for all effective curves $C$.  Treated over the rational
numbers $\Q$, this gives the {\it Zariski decomposition} 
\cite{Zariski}
\begin{equation}
D =  \sum_{i} \gamma_i C_i + X, \; \; \gamma_i \in \Q \,.
\label{eq:Zariski}
\end{equation}
This decomposition determines the minimal degree of vanishing of a
section of a line bundle over curves $C_i$ in the base.  For example,
on $\F_{12}$ we have an irreducible effective divisor $S$ with $S
\cdot S = -12, - K \cdot S = -10.$ Thus, $- K$ has a Zariski
decomposition $- K = (5/6) C + X$.  It follows that $-4K = (10/3) C +
X, -6K = 10 C + X$.  Since $f, g$ are sections of ${\cal O} (-4K), 
{\cal O} (-6K)$
respectively,  $f$ must vanish to degree 4 (= $\lceil 10/3\rceil$) on
$S$, and $g$ must vanish to degree 5 on $C$, implying that there is an
$\ge_8$ type singularity associated with the generic elliptic
fibration over $\F_{12}$.  This matches the well-known fact that the
gauge group of the generic F-theory model on $\F_{12}$ is $E_8$
\cite{Morrison-Vafa-I}.  This general principle was used in the
classification of all non-Higgsable clusters in \cite{clusters}, and
will be used as a basic tool throughout this paper.  Note that the
Zariski decomposition (\ref{eq:Zariski}) determines the {\it
  minimal} degree of singularity of $f, g$ over a given curve, but the
actual degree of vanishing can be  made
higher for specific models by tuning
the coefficients in the Weierstrass representation.

\subsection{Bases $B_2$ for EFS Calabi-Yau threefolds with large
  $h^{2, 1}$}
\label{sec:bases}

The bases $B_2$ that can support an elliptically fibered Calabi-Yau
threefold are complex surfaces, which can be characterized by the
structure of effective divisors (complex curves) on the surface.
Divisors on $B_2$ are formal integral linear combinations of algebraic
curves, which map to homology classes in $H_2 (B_2,\Z)$.  The
effective divisors are those where the expansion in terms of algebraic
curves has nonnegative coefficients; the effective divisors generate a
cone (the Mori cone, dual to the K\"ahler cone on cohomology classes)
in $H_2 (B_2,\Z)$.

As summarized in \S\ref{sec:intro},  the minimal model program for
classification of complex surfaces  and the results of Grassi
show that the only bases $B_2$ that can support an elliptically
fibered Calabi-Yau threefold
are $\P^2, \F_m (0 \leq m \leq 12),$ the Enriques surface,  and
blow-ups of these spaces.  The values of $h^{2, 1}$ for the generic
elliptic fibration over each of these surfaces can be read off from
the intersection structure of  each base using Table~\ref{t:clusters}
and
equations (\ref{eq:11}) and (\ref{eq:21}).
The intersection structure of divisors
on the bases $\F_m$ is quite simple.
$\F_m$ is a $\P^1$ bundle over $\P^1$, with $h^{1, 1}(\F_m) = 2$, so
$T = 1$.  The cone of effective divisor
classes
on each of these surfaces  is generated by $S, F,$ where $S$ is a
section of the $\P^1$ bundle with $S \cdot S = - m$, and $F$ is a
fiber with $F \cdot F = 0, F \cdot S = 1$.\footnote{Really,
$[F]$ is a class in $H_2$, and the fibers are  a continuous family of
  divisors in this class that foliate the total space; as mentioned
  earlier, we will generally go back and forth freely between divisors
and their associated classes.}
 
The $-12$ curve on $\F_{12}$ carries an $E_8$ gauge group, so the
generic elliptic fibration over this base has $r = 8, V= 248$ and
$h^{1, 1} = 11, h^{2, 1}= 491$.  Similarly, for $\F_{8}$ and $\F_7$ we
have $h^{1, 1} = 10, h^{2, 1}=376$, and for $\F_6$, $h^{1, 1} = 11, h^{2, 1}= 321$, with decreasing
values of $h^{2, 1}$ for $\F_m, m < 6$ (see \cite{toric} for a
complete list).  Since tuning Weierstrass coefficients to increase the
size of the gauge group or blow up points in the base entails a
reduction in $h^{2, 1}$, to construct all EFS CY3s with $h^{2, 1} \geq
350$, we need only consider the minimal bases $\F_{12}, \F_8$, and
$\F_7$.  Note that, as discussed,  for example,
in \cite{clusters}, $\F_m$ for $m =
9, 10, 11$ contain points on the $- m$ curve where $f, g$ must vanish
to degrees 4, 6, which must be blown up leading to a new base of the
form of $\F_{12}$ or a blow-up thereof, so the Hirzebruch surfaces
$\F_9, \F_{10}, \F_{11}$ are not good bases for an EFS CY3.

The irreducible effective divisors on $\F_m$ are those of the form $D
= a S + b F, b \geq m a$, since if $b < m a,$ then $D \cdot S< 0$ and
$D$ contains $S$ as a component (and is therefore reducible).  Blowing
up a base $B_2 = \F_m$ at a point $p$ produces a new $-1$ curve, the
{\it exceptional divisor} $E$ of the blow-up.  Each curve $C$ in $B_2$
that passes once smoothly through $p$ gives a {\it proper transform}
${C}'\sim C - E,$ with $E \cdot {C}'= 1$.  Since $\F_m$ is a $\P^1$
bundle over $\P^1$, each $p \in \F_m$ lies on some fiber in the class
of $F$.

\begin{figure}
\begin{center}
\includegraphics[width=12cm]{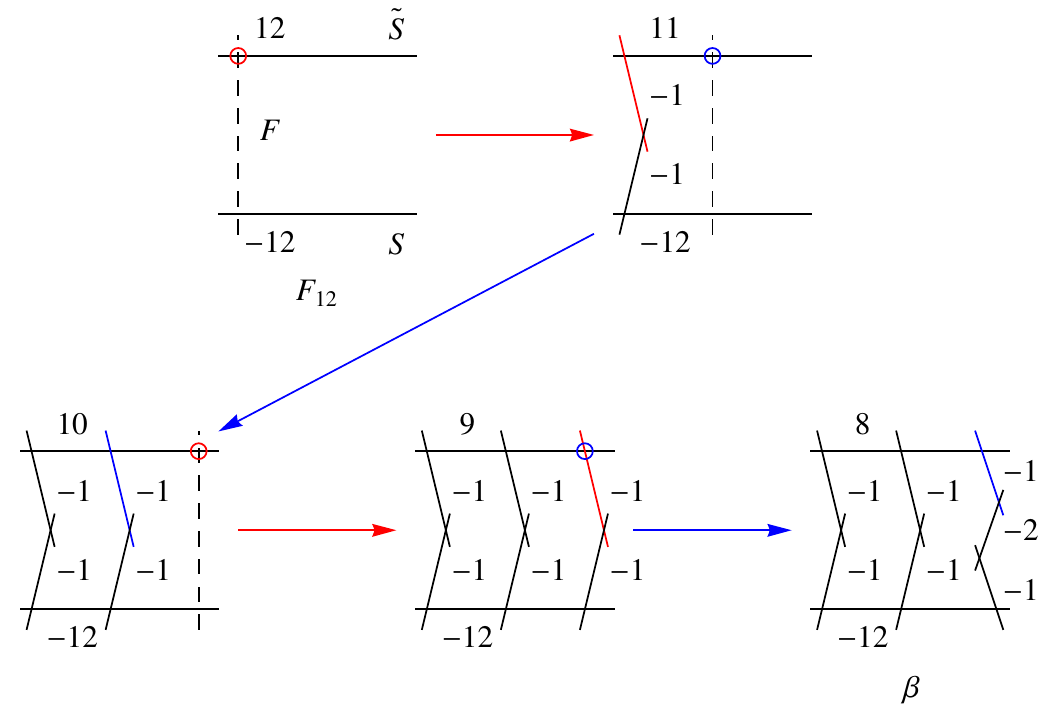}
\end{center}
\caption[x]{\footnotesize  A general F-theory base $B_2$ is formed by
  a sequence of blow-ups on a Hirzebruch surface $\F_m$.  In this
  example, three generic points are blown up sequentially on
  $\F_{12}$, and a fourth  blow-up point is chosen to be on the
  exceptional divisor from the third blow-up.  These points are all
  blown up on fibers in such a way that a global $\C^*$
structure is preserved.
The final base $\beta$ enters the discussion in the text in several places.}
\label{f:blow-up}
\end{figure}

We can describe a sequence of blow-ups on $\F_m$ by tracking the cone
of effective divisor classes after each blow-up.  The result of a single
blow-up at a generic point on $\F_m$ gives a new base $B_2',$ with an
exceptional divisor $E$ having $E \cdot E = -1$ extending the cone of
effective divisors in a new direction.  If we denote the specific
fiber of $\F_m$ containing $p$ as $F_1$, then $F_1' \sim F_1 - E$ is
also in the new cone of effective divisors, with $F_1' \cdot E = 1$.
There is also an effective divisor in the class $\tilde{S}= S + m F$
(with $\tilde{S}\cdot \tilde{S}= + m$) that passes through the generic
point $p$, and this gives a new curve $\tilde{S}'$ in $B_2'$ with
$\tilde{S}' \cdot \tilde{S}' = m -1$.  In this way, we can
sequentially blow up points on $\F_m$ to achieve any allowable base
$B_2$ for an EFS Calabi-Yau threefold.
An example of a sequence of bases formed from four
consecutive blow-ups of $\F_{12}$ is shown
in Figure~\ref{f:blow-up}.

A  point in the base must be blown up whenever
there is a $(4, 6)$
vanishing of $f, g$ at that point.  In general, such a singularity can
be arranged at a point in the base by tuning 29 parameters in the
Weierstrass model  \cite{Witten-phase-mf}.  This matches with the
gravitational anomaly cancellation condition $H-V= 273-29T$ (see
(\ref{eq:hv})), since a single new tensor field arises when the point
in base is blown up.    From (\ref{eq:11}) and (\ref{eq:21}) we thus see that,
generically, blowing up a point will cause a change in the Hodge
numbers of a base by
\begin{equation}
\Delta h^{1, 1} = + 1, \; \;
\Delta h^{2, 1} = -29 \,.
% \label{eq:}
\end{equation}
As an example, the final  base $\beta$ depicted in Figure~\ref{f:blow-up} is
associated with four blow-ups of $\F_{12}$, and thus has Hodge numbers
$h^{1, 1}= 11 + 4 = 15,h^{2, 1} = 491-4 \times 29 = 375$.
In some situations, when there is a gauge group
involved along divisors containing the blow-up point, there is also a
change in $V$ that modifies the number of moduli removed by the
blow-up, and correspondingly affects the Hodge numbers of the new
Calabi-Yau threefold.

In general, the combinatorial structure of the cone of effective
divisors on $B_2$ can become quite complicated.  A simple subclass of
the set of bases that are formed when multiple points on $\F_m$ are
blown up consists of those bases where the points blown up lie on $\phi$ distinct
fibers, and those blown up on each fiber are at the intersection of
irreducible effective divisors of negative self intersection lying
within that fiber or intersections between such divisors and the
sections $S, \tilde{S}$ of the original $\F_m$.  In this case, a
global $\C^*$-structure is preserved on the base $B_2$; bases of this
type were classified in \cite{Martini-WT}.  When $\phi \leq 2$, so
that all points blown up lie on two or fewer fibers, the base is
toric; the set of toric bases was classified in \cite{toric}.  In
cases where the number of fibers blown up satisfies $\phi \leq m$, the
initial point $p_i$ blown up on each fiber can be a generic point and
a representative of the class $\tilde{S}$ can be found that passes
through all these points, so that the base has a global $\C^*$
structure.  Almost all the bases we consider in this paper will have
this structure, and can be represented as $\C^*$-bases with $\phi$
nontrivial fibers.  We will discuss particular situations where we
need to go beyond this framework as they arise.

For the toric and $\C^{*}$-bases, an explicit representation of
the monomials in the Weierstrass model can easily be given, as described in
\cite{toric, Martini-WT}.  This representation is useful for explicit
calculations, as discussed further below in \S\ref{sec:Weierstrass}.

One issue that must be addressed in enumerating distinct bases for EFS
CY3s is the role of $-2$ curves in the base.  In general, isolated -2
curves, or connected clusters of $-2$ curves that do not carry a gauge
group, are realized at specific points in the moduli space of
fibrations over bases without those $-2$ curves.  For example, blowing
up $\F_{12}$ at two distinct generic points $p_1, p_2$ gives rise to two
nontrivial fibers, each containing two connected curves of
self-intersection (-1, -1) (like the left two fibers in the base
$\beta$ from Figure~\ref{f:blow-up}).  In the limit where $p_2$
approaches $p_1$, this becomes two blow-ups on a single fiber,
containing three connected curves of self-intersection (-1, -2, -1)
({\it e.g.,} the right-most fiber in Figure~\ref{f:blow-up}).  This
can be seen in the toric and $\C^*$ cases directly through the
enumeration of monomials, as discussed in \cite{toric, Martini-WT};
the $-2$ curves in clusters not associated with Kodaira singularities
giving nonabelian gauge groups correspond to extra elements of $h^{2,
  1}$ not visible in the explicit monomial count, and the
corresponding Calabi-Yau is most effectively described by the more
generic base where the blow-up points are kept distinct.  On the other
hand, when a $-2$ curve supports a nontrivial gauge group either due to
an NHC or a tuning, this curve is ``held in place'' by the singularity
structure, which would not be possible in the given form without the
$-2$ curve.  Thus, when enumerating all distinct possible EFS CY3s, we
should only include $-2$ curves in bases where $(f,g, \Delta)$ have
nonzero vanishing degrees over these curves\footnote{Note that there
  is one additional subtlety, which arises when a configuration of
  $-2$ curves describes a degenerate elliptic fiber \cite{Martini-WT},
  but this situation does not arise for any bases considered in this
  paper}.

By following these principles, we can systematically enumerate the
bases associated with EFS CY3s with large $h^{2, 1}$.
In almost all cases, the bases have a $\C^*$ structure and can be
described as $\F_{12}$ blown up at a sequence of points
along one or more fibers.  The precise sequences of possible blow-ups
are detailed in Section \ref{sec:systematic}.

\subsection{Constraints on codimension one singularities and
  associated gauge groups}
\label{sec:tuning} 

In this and the following sections, we describe in more detail
how codimension one
and two singularities in the
elliptic fibration of the Calabi-Yau threefold $X$ over a given base
$B_2$ can be understood and classified.  In this analysis we use the
physical language of F-theory; though in principle the arguments here
could be understood purely mathematically without reference to gauge
groups or matter, the physical F-theory picture is extremely helpful
in clarifying the geometric structures involved.

As we have described already, the NHCs of intersecting irreducible
effective divisors of negative self-intersection tabulated in
Table~\ref{t:clusters} give rise to nonabelian gauge groups and, in
some cases, charged matter over any base $B_2$ that contains these
clusters.  These physical features of the EFS CY3s encode the
topological structure of $X$ through equations (\ref{eq:11}) and
(\ref{eq:21}).  Additional and/or enhanced gauge groups and matter can
also be realized, giving rise to a range of different EFS CY3s over a
given base $B_2$, by tuning the parameters in the Weierstrass model
(\ref{eq:Weierstrass-1}).  
Over simple bases like $\P^2$, the
range of possible tunings is enormous, giving rise to many thousands
of topologically distinct CY3s elliptically fibered over the fixed base \cite{kpt, Braun-0}.
For the CY3s with large $h^{2, 1}$ that we consider here, however, the
range of possible tunings over the relevant bases $B_2$ is quite small.  

Some general constraints on when codimension one singularities can be
tuned beyond the minimal values required on a given base follow from
the Zariski decompositions of $-4K$ and $-6K$.
These constraints provide strong bounds on the set of possible gauge
groups that can be tuned over any given $B_2$.  These constraints,
which we analyze in general terms in this section, do
not, however, guarantee the existence of a given tuned model with
specific gauge groups.  To confirm that a Weierstrass model can be
realized, a more detailed analysis is needed, as discussed in the
subsequent sections.

Consider a rational curve\footnote{A rational curve is a complex curve
  of genus 0; it is shown in \cite{clusters} that an effective divisor
  in the base of an elliptically fibered Calabi threefold cannot be a
  higher genus curve of negative self-intersection without forcing a
  (4, 6) vanishing of $f, g$.} $C$ of self-intersection $C \cdot C = -
k$.  From $(K + C) \cdot C = 2g-2 = -2$, we have $K \cdot C = k-2$.
Consider a divisor $D = - nK$ that contains as irreducible
components a set of curves $B_i$ with multiplicities $b_i$ that each
intersect $C$
simply at a single point: $B_i \cdot C = 1$.  Then we have
\begin{equation}
 D = c C + \sum_{i} b_i B_i + X, \;\;\;\;\; {\rm with} \; X \cdot C
 \geq 0 \,,
% \label{eq:}
\end{equation}
where $D \cdot C = - nK \cdot C = - n(k -2),$ so a section of (the
line bundle associated with) $D$ must vanish at least $c$ times on
$C$, where
\begin{eqnarray}
 X \cdot C & =& (D - c C - \sum_{i}b_i B_i) \cdot C
= - n(k -2) + k c - \sum_{i} b_i \geq 0.\\
\Rightarrow & &  c \geq \frac{1}{k} (\sum_{i}b_i + n(k -2))
\label{eq:c-relation}\,.
\end{eqnarray}
This result has a number of specific consequences for where
codimension one singularities can be tuned on bases with a given
configuration of non-Higgsable clusters from Table~\ref{t:clusters},
which are connected in any given base by a network of $-1$ curves.
We give some specific examples:
\vspace*{0.05in}

\noindent
{\bf  No $(f, g)$ tuning 
can give a nonabelian gauge algebra
on $-1$ curves connected to any singular
  cluster other than a single $-3$ curve.}

Consider, for example, a $-4$ curve $C$ that intersects a $-1$ curve $B$.
The minimal $(f, g)$ tuning on $B$ needed to get a nontrivial Kodaira
singularity ({\it i.e.,} one which gives rise to a nonabelian gauge
algebra) is (1, 2).
Applying (\ref{eq:c-relation}) with $k = 4$ for $n= 4, b_1 = 1$ gives
$c_{(4)}\geq 9/4$, and for $n= 6, b_1 = 2$ gives $c_{(6)}\geq 13/4$,
so tuning a (1, 2) vanishing on a $-1$ curve
$B$ that intersects  a $-4$  curve $C$ forces a (3, 4) vanishing on $C$,
which means that $f,  g$ vanish to degrees (4, 6) at the point $B
\cdot C$, which cannot happen on a good base $B_2$ for an EFS CY3.
A similar argument shows that $(f, g)$ cannot be tuned to vanish to
degrees (1, 2) on a $-1$ curve that intersects any of the other NHCs
that carry a nontrivial gauge group other than one or two isolated -3
curves.  A $-1$ curve that intersects a $-3$ curve can carry an $(f, g)$
vanishing of (1, 2), while the $-3$ curve carries a (2, 3) vanishing.
Note that a $-1$ curve $C$ intersecting three $-3$ curves each with
vanishing (2, 3) would have by (\ref{eq:c-relation}) $c_{(4)}\geq 2,
c_{(6)}\geq 3,$ so $C \cdot B_i$ would correspond to points of (4,
6) vanishing.
\vspace*{0.05in}

\noindent
{\bf Vanishing of $f, g, \Delta$ on a $-2$ curve} 

From (\ref{eq:c-relation}), the degrees of vanishing of $f, g$, or $\Delta$ on
any $-2$ curve $C$ must be
\begin{equation}
 c \geq \sum_{i} \frac{b_i}{2}
% \label{eq:}
\end{equation}
where $b_i$ are the degrees of vanishing of $f, g$, or $\Delta$ on curves $B_i$
that intersect $C$. 
We refer 
to this rule for degrees of vanishing on $-2$ curves as the
``averaging rule'' in later arguments, where it will be of use in
the analysis of toric and  $\C^*$
bases, for which each divisor in a fiber intersects
precisely two neighboring divisors.  
\vspace*{0.05in}

As an example, in the non-Higgsable (-3, -2) cluster, the degrees of
vanishing of $(f, g)$ on the $-3$ and $-2$ curves are, respectively, (2,
3), and (1, 2), which satisfy the above inequality ({\it e.g.,} for
$g$, $b_i = 3, c_{(6)}= 2 \geq 3/2$).
\vspace*{0.05in}

These rules, and applications of (\ref{eq:c-relation}) in a variety of
other cases, strongly constrain the places where extra codimension one
singularities can be tuned over EFS CY3s with large $h^{2, 1}$.  In
general, a tuning is only possible when sections of ${\cal O} (-4K), 
{\cal O} (-6K)$ can be
found in the form $D = \sum_{i} c_i C_i + X$ with no divisor or points
where $(f, g)$ vanish to degrees (4, 6).

\subsection{Anomalies and matter content}
\label{sec:anomalies}

Another set of geometric constraints are encoded in the detailed anomaly
cancellation equations of 6D supergravity theories.  
For an F-theory compactification on a base $B_2$ with canonical class
$K$ and nonabelian gauge group factors $G_i$ associated with
codimension one singularities on divisors $S_i$, the anomaly
cancellation conditions are
\cite{gswest, Sagnotti, Erler, Sadov, KMT-II}
\begin{eqnarray}
H-V & = &  273-29T\label{eq:hv}\\
0 & = &     B^i_{\rm adj} - \sum_{\bf R}
x^i_{\bf R} B^i_{\bf R} \label{eq:f4-condition}\\
K \cdot  K & =   &9 - T  \label{eq:aa-condition}\\
-K \cdot S_i & =  & \frac{1}{6} \lambda_i  \left(  \sum_{\bf R}
x^i_{\bf R} A^i_{\bf R}-
A^i_{\rm adj} \right)  \label{eq:ab-condition}\\
S_i\cdot  S_i & =  &\frac{1}{3} \lambda_i^2 \left(  \sum_{\bf R} x_{\bf
  R}^i C^i_{\bf R}  -C^i_{\rm adj}\right)  \label{eq:bb-condition}\\
S_i \cdot S_j & = &  \lambda_i \lambda_j \sum_{\bf R S} x_{\bf R S}^{ij} A_{\bf R}^i
A_{\bf S}^j\label{eq:bij-condition}
\end{eqnarray}
where  $A_{\bf R},
B_{\bf R}, C_{\bf R}$ are group theory coefficients defined through
\begin{align}
\tr_{\bf R} F^2 & = A_{\bf R}  \tr F^2 \\
\tr_{\bf R} F^4 & = B_{\bf R} \tr F^4+C_{\bf R} (\tr F^2)^2 \label{eq:bc-definition}\,,
\end{align}
$\lambda_i$ are numerical constants associated with the different
types of gauge group factors ($\lambda = 1$ for $SU(N)$, 2 for $SO(N)$
and $G_2$),
and where
$x_{\bf R}^i$ and $x_{\bf R S}^{ij}$
denote the number of matter fields that transform in each irreducible
representation ${\bf R}$ of the gauge group factor $G_i$
and $({\bf R} , {\bf S})$ of $G_i \otimes G_j$ respectively. (The
unadorned ``tr'' above denotes a trace in the fundamental
representation.) 
Note that for groups such as $SU(2)$ and $SU(3)$, which lack a fourth
order invariant, $B_{\bf R} = 0$ and there is no condition
\eq{eq:f4-condition}.
The group theory coefficients for the representations relevant for
this paper are compiled for convenience in Table~\ref{t:coefficients}

\begin{table}
\begin{center}
\begin{tabular}{||c |c || c |c |c  ||}
\hline
Group & Rep & $A_{\bf R}$ & $B_{\bf R}$ & $C_{\bf R} $\\
\hline
$SU(2)$ & {\bf 2} & 1 &---&$\frac{1}{2}$\\
& {\bf 3} & 4 &---& 8\\
\hline
$SU(3)$ & {\bf 3} & 1 &---&$\frac{1}{2}$\\
& {\bf 8} & 6 &---& 9\\
\hline
$G_2$ & {\bf 7} & 1 &---& $\frac{1}{4}$\\
& {\bf 14} & 4 &---&$\frac{5}{2}$ \\
\hline
\end{tabular}
\end{center}
\caption[x]{\footnotesize Group theory coefficients $A_{\bf R}, B_{\bf
    R}, C_{\bf R}$ for fundamental and adjoint matter representations
  of gauge groups relevant for the analysis of this paper.
  Note that the gauge groups $SU(2), SU(3), G_2$ have no fourth order
  Casimir so there are no coefficients $B_{\bf R}$.}
\label{t:coefficients}
\end{table}

The 6D anomaly cancellation conditions provide additional constraints
on the set of possible structures for EFS Calabi-Yau threefolds.  For
any set of possible gauge groups satisfying the Zariski conditions
described in the previous section, the anomaly cancellation conditions
can be used to further check the consistency of the model and to
compute the possible matter spectra, giving $H_{\rm charged}$, which
can then be used in (\ref{eq:21}) to compute $h^{2, 1}$.  For example,
consider tuning a gauge group $SU(2)$ on a curve $C$
of genus $g$ and
self-intersection $-n$.
Assuming only fundamental ({\bf 2}) and adjoint ({\bf 3}) matter,
the spectrum of fields charged under this gauge group is uniquely
determined by the anomaly cancellation conditions
\begin{eqnarray}
K \cdot C  = 2g +n-2& = & \frac{1}{6} \left( A_{\bf 3} (1-x_{\bf 3})
-A_{\bf 2}x_{{\bf    2}}\right)  = 2/3-x_{\bf 2}/6  -2x_{\bf 3}/3\\
C \cdot C = -n & = &  \frac{1}{3}  \left( C_{\bf 3}(x_{\bf 3} -1)
 + C_{\bf 2}x_{\bf 2} \right) = 8(x_{\bf 3} -1)/3+ x_{{\bf 2}}/6  \,,
\end{eqnarray}
to be $x_{\bf 3} = g, x_{\bf 2} = 16-6n-16g$.  For a rational curve
$C$ with $g = 0$, there are simply $16-6n$ fields in the fundamental
({\bf 2}) representation.  This matches with the expectation that when
$-n \leq -3$ there is a larger gauge group and an $SU(2)$ is
impossible.  For higher genus curves $g$ the number of fields in the
adjoint is generically $g$ with no higher-dimensional matter
representations.  For specially tuned models, higher matter
representations are possible, but for $\gsu (2)$
all representations other than ${\bf
  2}$ contribute to the genus \cite{kpt, mt-singularities}.  Gauge
groups on higher genus curves and associated exotic matter
representations of this type do not appear in the models considered
here at large $h^{2, 1}$, and are discussed further in
\S\ref{sec:issues-matter}.

From the gauge group and matter content associated with a given tuned
Weierstrass model, the Hodge numbers can be computed from
(\ref{eq:11}), (\ref{eq:21}).  Continuing with the preceding example,
tuning an $SU(2)$ gauge group on a divisor of self-intersection $-n$
that does not intersect any other curves carrying gauge groups leads
to a change in Hodge numbers of
\begin{eqnarray}
\Delta h^{1, 1}&= & \Delta r= + 1, \\
\Delta h^{2, 1}& = & \Delta V-\Delta H_{\rm charged} = + 3-2 (16-6n)
= -29 + 12n \,.
\end{eqnarray}
It is straightforward to compute the contribution to the Hodge numbers
from tuning any of the other gauge groups associated with a Kodaira
singularity type on a rational curve of given self-intersection.  
Table~\ref{t:matter-1} tabulates these values  for the gauge group
factors that are relevant for this paper.

\begin{table}
\begin{center}
\begin{tabular}{|c |c| c | c |}
\hline
& matter & $\Delta h^{1, 1}$ & $\Delta h^{2, 1}$\\
\hline
$\gsu (2) $ &  $(16-6n)$ $\times $ {\bf 2} &  $+1$ & $-29 + 12n$\\
$\gsu (3) $ & $(18-6n)$ $ \times $ {\bf 3}& $+ 2$ & $-46 + 18n$\\
$\gg_2 $ & $(10-3n) \times$ {\bf 7}& $+ 2$ & $-56 + 21n$\\
\hline
\end{tabular}
\end{center}
\caption[x]{\footnotesize Table of matter content and Hodge number
  shifts for tuned gauge algebra summands on a $-n$ curve $C$.  Shifts are
  computed assuming the curve carries no original gauge group; for $n
  \geq 3$ the contribution from the associated non-Higgsable cluster must be
  subtracted.  These shifts also do not include any necessary modifications for
  bifundamental matter, which must be taken into account when $C$
  intersects other curves carrying a gauge group.}
\label{t:matter-1}
\end{table}

Finally, the anomaly cancellation condition (\ref{eq:bij-condition})
indicates that when two curves $C, D$ intersect and both carry gauge
groups, a certain part of the matter is charged under both gauge group
factors.  This bi-charged matter is a subset of the total charged
matter content in each case, and must be taken into account when
computing the Hodge numbers of a threefold with this structure in the
base.  For example, two $SU(2)$ factors tuned on two intersecting $-2$
curves each have, from Table~\ref{t:matter-1}, 4 fundamental matter
fields.  From (\ref{eq:bij-condition}), there is one bifundamental
matter field transforming in the ${\bf 2} \times {\bf 2}$
representation.  This field, which contains 4 complex scalars, is
counted in the matter charged under each of the $SU(2)$.  Thus, while
the change in $h^{2, 1}$ from tuning each of these $SU(2)$ factors
individually is $\Delta h^{2, 1} = -5$, the net change from tuning
both of these factors is $-6$; {\it i.e.}, the second $SU(2)$ requires
tuning only a single additional Weierstrass modulus.

\subsection{Weierstrass models}
\label{sec:Weierstrass}

While the Zariski decomposition of $f, g, \Delta$, and the anomaly
cancellation conditions described in the last two sections place
strong constraints on the set of possible gauge groups and matter
fields that can be tuned in a Weierstrass model over any given base
$B_2$, these constraints are necessary but not sufficient for the
existence of a consistent geometry.  To prove that a given Calabi-Yau
geometry exists, it is helpful to consider an explicit construction of
the Weierstrass model.  This can be done in a straightforward way for
toric bases using the explicit realization of the monomials in the
Weierstrass model as elements of the lattice $N^*$ dual to the lattice
$N$ in which the toric fan is described.  This approach generalizes in
a simple way to bases that admit only a single $\C^*$ action.  The
details of this analysis are worked out in detail in \cite{toric,
  Martini-WT}.  It is also possible to
describe Weierstrass models explicitly for bases that are not toric or
$\C^*$, though there is at present no general method for doing this
and the analysis must be done on a case-by-case basis.  Explicit
construction of the monomials in a given Weierstrass model plays two
important roles in analyzing the Calabi-Yau threefolds we consider in
this paper.  First, by imposing the desired vanishing conditions for
$f, g, \Delta$ on all curves carrying gauge groups, we can check the
explicit Weierstrass model to confirm that no additional vanishing
conditions are forced on any curves or points that would produce
additional gauge groups or force a blow-up or invalidate the model due
to (4, 6) points or curves.  Second, we can perform an explicit check
on the value of $h^{2, 1}$ computed using the last term in
(\ref{eq:21}) by relating the number of free degrees of freedom in the
Weierstrass model to the number of neutral scalar fields.  This
analysis can, among other things, reveal the presence of additional
$U(1)$ gauge group factors that contribute to $V$ and $r$.  In
\cite{Martini-WT}, for example, it was found using this type of
analysis that a small subset of the possible $\C^*$-bases for EFS
Calabi-Yau threefolds give rise to generic nonzero Mordell-Weil rank.

We summarize here the relationship between $h^{2, 1}$ and the number
of Weierstrass monomials $W$ for a generic elliptic fibration over a
$\C^*$ base:
\begin{equation}
h^{2, 1}(X) = H_{\rm neutral} -1 =
W-w_{\rm aut} + N -4 + N_{-2} -G_1 \,,
\label{eq:monomials-w}
\end{equation}
where $w_{\rm aut} = 1 + {\rm max} (0, 1 + n_0, 1 + n_\infty)$ is the
number of automorphism symmetries, with $n_0, n_\infty$ the
self-intersections of the divisors coming from $S, \tilde{S}$, $N$ is
the number of fibers containing blow-ups, $N_{-2}$ is the number of
$-2$ curves that can be removed by moving to a generic point in the
moduli space of the associated threefold, and $G_1$
is the number of $-2$ curve combinations that represent a degenerate
elliptic fiber.  The relation (\ref{eq:monomials-w}) is a slight
refinement of the relation determined in \cite{Martini-WT} to include
tuned Weierstrass models; in particular, when considering tuned
(non-generic) elliptic fibrations over a given base the set of $-2$
curves contributing to $N_{-2}$ does not include certain $-2$ curves where
$\Delta$ vanishes to some order, even if this $-2$ curve is not in a
non-Higgsable cluster supporting a nonabelian gauge group.  We
encounter an example of this in the following section.

\subsubsection{Weierstrass models: some subtleties}
\label{sec:Weierstrass-subtleties}

As mentioned earlier, and also discussed in \cite{toric, Martini-WT},
curves of self-intersection $-2$ must be treated carefully when
analyzing the Weierstrass monomials and corresponding Hodge numbers.
$-2$ curves that do not carry vanishing degrees of $f, g, \Delta$ in
most circumstances are associated with special codimension one loci in
Calabi-Yau moduli space, and indicate additional elements of $h^{2,
  1}$ that are not visible in the Weierstrass monomials for the model
with the $-2$ curve.  To consistently distinguish different
topological types of Calabi-Yau threefolds, we should generally only
consider the most generic bases in each moduli space component, which
have no $-2$ curves on which $f, g, \Delta$ do not vanish to some
degree.  For example, the Weierstrass model describing the base
$\beta$ appearing in Figure~\ref{f:blow-up} has one fewer parameter
than expected for the given Calabi-Yau threefold, corresponding to a
contribution of $N_{-2} = 1$ in (\ref{eq:monomials-w}).  The generic base
for this threefold is given by blowing up four completely generic
points in $\F_{12}$, which gives four distinct $(-1, -1)$ fibers; the
base $\beta$ that contains a $-2$ curve in one fiber arises at limit
points of the moduli space where one of the blow-up points lies on the
exceptional divisor produced by one of the other blow-ups.  The
generic elliptic fibration over $\beta$ thus lives on the same moduli
space as the generic elliptic fibration over the base with four $(-1,
-1)$ fibers.  If, on the other hand, we tune an $SU(2)$ factor on the
top $-1$ curve of the $(-1, -2, -1)$ fiber, then the $-2$ curve
acquires a degree of vanishing of $\Delta$ of at least 1, and it is
fixed in place by the structure of the singularity.  This $SU(2)$
factor cannot be tuned in the bulk of the moduli space of the generic
four-times blown up $\F_{12}$.  In this situation,
$N_{-2} = 0$, and it can be checked that the $\C^*$
Weierstrass model contains the correct number of monomials.
  \footnote{The fact that additional structure can appear
  associated with $-2$ curves also arises in a related context
  in 4D heterotic theories based on elliptically fibered Calabi-Yau
  threefolds over bases containing these curves \cite{Anderson-WT}.}

Another subtlety that must be taken into account when computing the
number of free parameters for a Weierstrass model with given
codimension one singularity types is the appearance of each of the
gauge group factors $SU(2), SU(3)$ in two distinct ways in the Kodaira
classification.  In a generic situation, in the absence of other gauge
groups, an $SU(2)$ or $SU(3)$ gauge group tuned by a Kodaira type III
or IV singularity, as listed in Table~\ref{t:Kodaira} is simply a
special case of a type $I_2$ or $I_3$ singularity, and the complete
set of degrees of freedom needed to compute $h^{2, 1}$ should be
computed by imposing only the latter conditions.  In other cases,
however, such as in the context of non-Higgsable clusters, the type
III or IV singularity type may be forced by the structure of other
gauge groups or divisors.  In this case the specified gauge group
structure may not be possible with an $I_n$ singularity type, in which
case there are no monomials associated with such additional freedom.

Finally, for those gauge algebra types that depend not only on the
degrees of vanishing of $f, g, \Delta$, but also on monodromy, the
correct counting of degrees of freedom in the Weierstrass model
depends on the monodromy conditions.  The monodromy conditions for
each of the gauge group choices in type $IV, I_0^*,$ and $IV^*$
Kodaira singlets are described in \cite{Bershadsky-all, Morrison-sn},
and can easily be characterized in terms of the structure of monomials
in the Weierstrass model \cite{Anderson-WT}.

For all the models considered here, we have carried out an explicit
construction of the Weierstrass monomials, and confirmed that the
appropriate geometric structure exists and that the number of
monomials properly matches the value of $h^{2, 1}$, when the proper
shifts according to $-2$ curves and automorphisms
as described in (\ref{eq:monomials-w}) are taken into account.
For all the models considered here, the blow-ups on the different
fibers are independent, since the gauge groups on $S, \tilde{S}$ do
not change.  This means that the monomial analysis can be performed in
a local chart around each fiber independently, without loss of generality.

\subsubsection{Constraints on Weierstrass models: an example}

As an example of the utility of the explicit Weierstrass monomial
construction, we consider a simple example of a situation in which the
Zariski and anomaly analyses suggest that a tuning may be possible,
but it is ruled out by explicit consideration of the Weierstrass
model.

\begin{figure}
\begin{center}
\includegraphics[width=8cm]{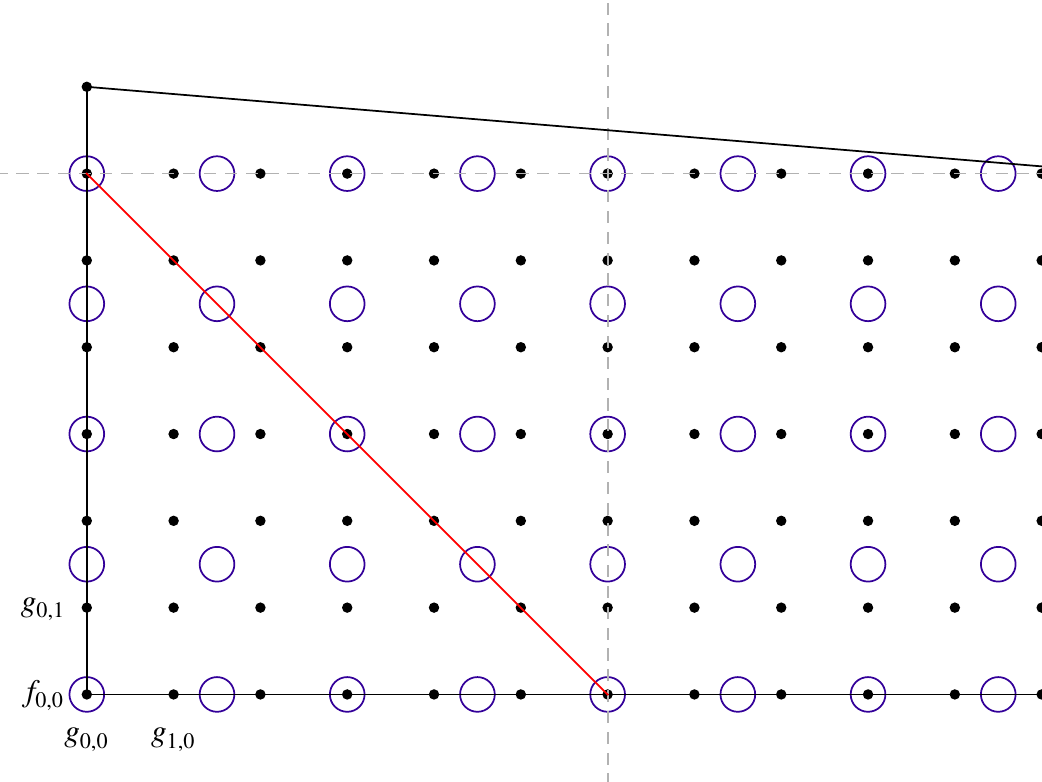}
\end{center}
\caption[x]{\footnotesize  Monomials in the generic Weierstrass model
  over $\F_{12}$ are of the form $f_{k, m}z^kw^m, g_{k, m}z^kw^m$, and
can be associated with points depicted above
in the  lattice $N^*$
dual to the lattice $N$ carrying the rays in the toric fan for
$\F_{12}$. Circles denote monomials in $f$, and dots denote monomials
in $g$.
Blowing up a generic point in $\F_{12}$ can be described
in a local coordinate system by setting all monomials below the red
line to vanish.  As described in the text, an $SU(2)$ gauge group
cannot be tuned on the exceptional divisor from the blow-up without
forcing the monomial coefficient $g_{0, 7}$ to vanish, which makes it
impossible to form a Calabi-Yau due to a (4, 6) vanishing on the
divisor
$S$.}
\label{f:example-monomials}
\end{figure}

Consider again the base $B_2 = \beta$ depicted in Figure~\ref{f:blow-up}.  
We can ask if an $SU(2)$ can be tuned through an $I_2$
$(0, 0, 2)$ singularity on the top $-1$ curve $C$
of one of the $(-1, -1)$
fibers.  (In fact, this analysis is equivalent for any such fiber on
$\F_{12}$, since as discussed above the analysis is essentially local
on each fiber in this situation where there is no change in the degree
of vanishing of $f, g, \Delta$ on $S, \tilde{S}$.)  
The $SU(2)$ that we might tune in this fashion does not violate any
conditions visible from the Zariski analysis, since we can take
$\Delta= 2C + 10S+ X$, and still satisfy $X \cdot D = 0$ where $D$ is
the lower $-1$ curve connecting $C$ and $S$.  
(Note, however, that we cannot have a type $III$ or $IV$ $SU(2)$ on
$C$, since this would force a vanishing of $\Delta$ on $D$.)
Tuning an $SU(2)$ on $C$ also does not present
any problems involving anomalies, since we have sufficient
hypermultiplets to have an $SU(2)$ with the requisite 10 fundamental
matter fields.  This configuration is, however, ruled out by an
explicit Weierstrass analysis.  In the toric language \cite{Fulton,
  toric}, we can take $\F_{12}$ to have a toric fan given by vectors
$v_i \in
N =\Z^2$:
$v_1 = (0, 1), v_2 = (1, 0), v_3 = (0, -1), v_4 = (-1, -12)$.  The allowed Weierstrass monomials for the generic elliptic
fibration over $\F_{12}$ are then $u \in N^*, \langle u, v_i \rangle
\geq -n$ with $n = 4, 6$ for $f, g$ respectively.  Taking a local
coordinate system where $z = 0$ on the fiber $F$
associated with $v_2$, and $w = 0$ on $\tilde{S}$, the
allowed monomials in $f = f_{k, m}z^k w^m$,
$g = g_{k, m}z^kw^m$
are those with $k, m \geq 0$, $12(m -n) + (k -n) \leq n$; these
degrees of freedom 
are depicted in Figure~\ref{f:example-monomials}.  The only monomial
that keeps $S$ from having a $(4, 6)$ singularity is the $w^7$ term in
$g$, so the coefficient $g_{0, 7}$ cannot vanish without breaking the
Calabi-Yau structure.  Blowing up the point of intersection between
$F$ and $\tilde{S}$ 
adds the vector $v_5=(1,1)$ to the toric fan, so we must remove the
monomials $u$ with $\left\langle  u, v_5\right\rangle < -n$ from $f,
  g$; in the chosen coordinates, this amounts to removing all
  monomials such that $m+k < n$, as 
depicted by the red diagonal line in the figure.  
With a change of coordinates $z = \zeta x, w = x$, $f = \hat{f}x^4$, 
$g = \hat{g}x^6$, we have a local expansion around $E, F' = F-E$
with coordinate $x =
0$ on $E$.  We can then expand
\begin{eqnarray}
\hat{f} (\zeta, x) & = &   \hat{f}_0 (\zeta) + \hat{f}_1 (\zeta) x + \cdots\\
& = &
(\hat{f}_{0, 0} + \hat{f}_{1, 0} \zeta + \cdots \hat{f}_{4, 0} \zeta^4)
+ (\hat{f}_{1, 1} \zeta + \hat{f}_{1, 1} \zeta^2 + \cdots \hat{f}_{5, 1} \zeta^5) x + \cdots\\
\hat{g} (\zeta, x) & = &   \hat{g}_0 (\zeta) + \hat{g}_1 (\zeta) x + \cdots \\
& = &
(\hat{g}_{0, 0} + \hat{g}_{1, 0} \zeta + \cdots \hat{g}_{6, 0} \zeta^6)
+ (\hat{g}_{0, 1} + \hat{g}_{1, 1} \zeta + \cdots \hat{g}_{7, 1} \zeta^7) x + \cdots
\end{eqnarray}
The condition that $\Delta$ vanish at order $x^0$ requires that $4\hat{f}_0^3 + 27\hat{g}_0^2 = 0$, which we can satisfy by
setting $\hat{f}_0 (\zeta) =- 3 \alpha^2, \hat{g}_0 (\zeta) = 2
\alpha^3$ for some quadratic function $\alpha (\zeta)$.  The condition that
$\Delta$ vanish at order $x$ then requires that
\begin{equation}
2\hat{f}_0^2 \hat{f}_1 +  9\hat{g}_0\hat{g}_1 = 0 \,.
\label{eq:linear-condition}
\end{equation}
This condition cannot, however, be satisfied when $\alpha \neq 0$,
without setting $\hat{g}_{0, 1} = 0$, since $\hat{f}_1$ contains no
term of order $\zeta^0$.  But $\hat{g}_{0, 1}= g_{0, 7} = 0$ forces
$g$ to vanish to 
degree 6 on $S$ so there would be a degree (4, 6) singularity on $S$,
which is incompatible with the Calabi-Yau structure.
Thus, we cannot tune an $I_2$ $SU(2)$ singularity on $C$.
Note that while the coordinates $\zeta, x$ make this computation
particularly transparent, the same result can be derived directly in
the $z, w$ coordinates.  In particular, this means that an $SU(2)$
cannot be tuned on the curve in question even if further points on the
base are blown up.

Note also that while this analysis rules out an $SU(2)$ on the $-1$ curve $C$
in question, it  is still possible to tune $\Delta$ to vanish to
second order on this curve.  If $\hat{f}_0 = \hat{g}_0 = 0$, then
(\ref{eq:linear-condition}) is automatically satisfied.  This allows
for the possibility of a $(1, 1, 2)$ vanishing of $(f, g, \Delta)$ on
$C$. Indeed, such a vanishing -- which does not lead to any gauge
group -- arises in some configurations for EFS CY threefolds, as we
see below.

This kind of analysis can be used to check explicitly
whether a  Weierstrass model exists for any given combination of gauge
group tunings that satisfy the Zariski and anomaly cancellation
conditions.  This is straightforward for the gauge groups that are
imposed by particular orders of vanishing of $f, g$, since this
corresponds simply to setting the coefficients of certain monomials in
these functions to vanish.  The analysis is more subtle, however, for
type $I_n$ and $I_n^*$ singularities, such as the $I_2$ example
considered here, where vanishing on $\Delta$ requires more complicated
polynomial conditions on the coefficients.  
For large $n$, the algebra involved in explicitly imposing an $I_n$
singularity can be quite involved.
This is not an issue for
any of the threefolds considered in this paper, but presents a
technical obstacle to a systematic analysis for general $h^{2, 1}$.
We return to this issue in \S\ref{sec:issues-classical}.

Finally, note that the fact that an $SU(2)$ cannot be tuned on the top
$-1$ curve of a $(-1, -1)$ fiber matches with the example described in
\S\ref{sec:Weierstrass-subtleties}, where an $SU(2)$ tuned on the top
curve of a $(-1, -2, -1)$ fiber fixes the middle $(-2)$ curve in
place.  The lower $-1$ curve cannot be moved to a different location
on the $-12$ curve $S$, which would remove the $-2$ curve, since this
would leave behind precisely the configuration we have just ruled
out.  This confirms that this $-2$ curve does not represent a missing
modulus and does not contribute to $N_{-2}$ in (\ref{eq:monomials-w}),
even though it does not itself support a gauge group.

\section{Systematic construction of EFS CY threefolds with large
$h^{2, 1}$} 
\label{sec:systematic}

\begin{figure}
\begin{center}
\includegraphics[width=11cm]{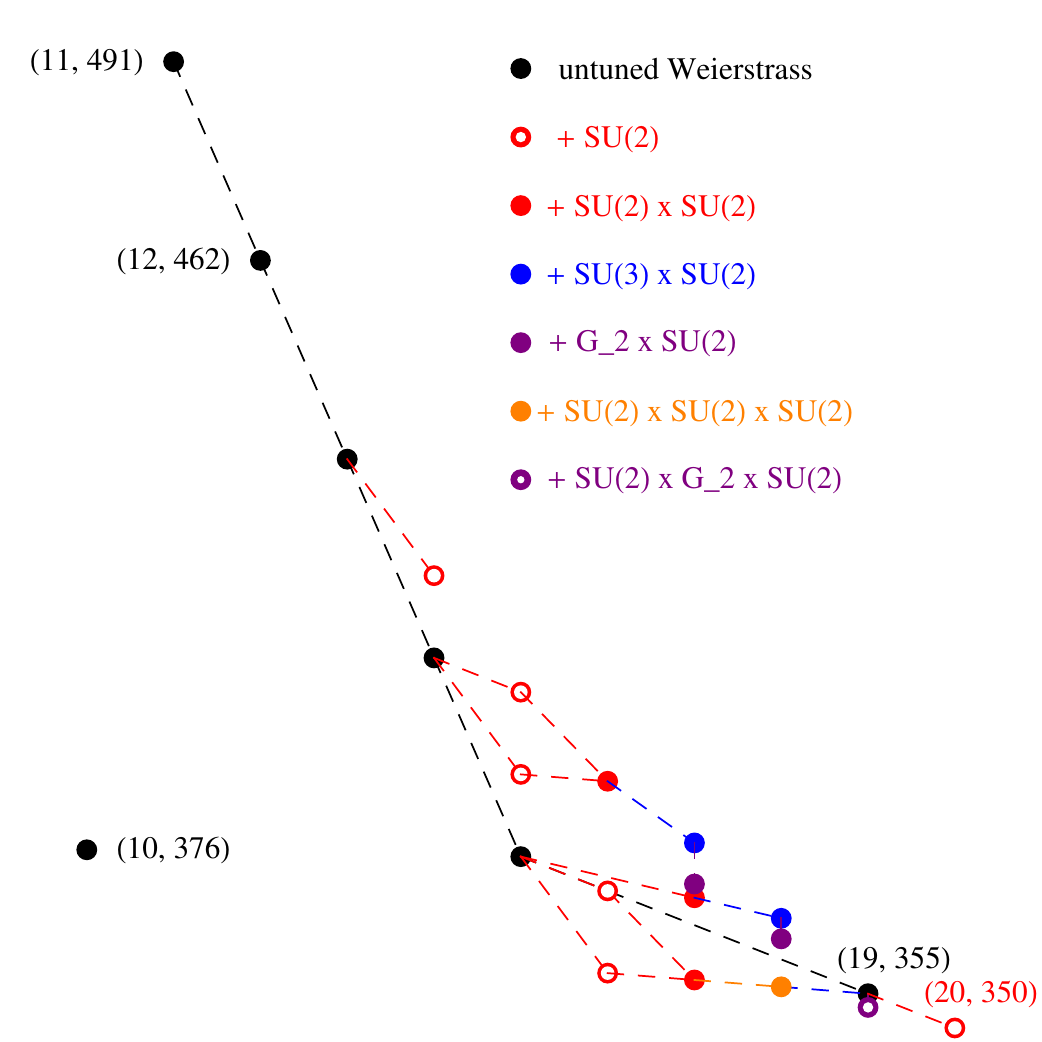}
\end{center}
\caption[x]{\footnotesize  
In this paper we explicitly construct all
elliptically fibered Calabi-Yau threefolds with
section having $h^{2, 1}\geq 350$.  The Hodge numbers of these
threefolds are shown here, with the detailed construction explained in
the bulk of the text.  Black points represent generic
elliptic fibrations over different bases $B_2$, and colored points
represent tuned Weierstrass models over these bases with enhanced
gauge groups.  The three purple data points appear to be new Calabi-Yau
manifolds not found in the Kreuzer-Skarke database (see
\S\ref{sec:new}).  All elliptically 
fibered Calabi-Yau threefolds with section are connected by geometric
transitions associated with tuning Weierstrass moduli over a
particular base (``Higgsing/unHiggsing'') and/or blowing up and down
points in the base (corresponding to tensionless string
transitions in the physical F-theory context).  Note that the point $(10,
376)$, corresponding to generic elliptic fibrations over $\F_7, \F_8$,
is connected to the other threefolds shown through a sequence of
blow-up and blow-down transitions on the base that pass through
the set of threefolds with smaller Hodge numbers $h^{2, 1}< 350$. 
Note also that there are two distinct constructions that give the
Hodge numbers $(19, 355)$; in addition to an untuned Weierstrass model
with generic gauge group $G_2 \times SU(2)$ there is a tuning of the generic $(15,
375)$ Weierstrass model with a gauge group $SU(2) \times SU(3) \times SU(2)$.}
\label{f:results}
\end{figure}

We now systematically describe how all Calabi-Yau threefolds that are
elliptically fibered with section (EFS) and have $h^{2, 1}\geq 350$
are constructed by tuning gauge groups on $\F_{12}, \F_8, \F_7,$ and
blow-ups thereof.  We begin with the Hirzebruch surfaces and consider
all possible tunings that would give a threefold with $h^{2, 1} \geq
350$.  For those tunings that are possible by the Zariski and anomaly
cancellation conditions we check the Weierstrass models explicitly
using the toric monomial method.  For each set of valid Hodge numbers
we compare with the Kreuzer-Skarke database \cite{Kreuzer-Skarke} of
Hodge numbers for Calabi-Yau threefolds realized as hypersurfaces in
toric varieties using the Batyrev construction \cite{Batyrev}.  The
final results of our analysis are compiled in Figure~\ref{f:results},
and the full set of constructions is listed in
Table~\ref{t:Hodge-350}.

\subsection{Tuning models over $\F_{12}$}
\label{sec:tuning-0}

To systematically construct all Calabi-Yau threefolds that are
elliptically fibered with section, beginning with the largest value of
$h^{2, 1}$ and preceding downward, we begin with the generic elliptic
fibration over $\F_{12}$.  As described above and in \cite{WT-Hodge},
this Calabi-Yau threefold has Hodge numbers
$(h^{1, 1},h^{2, 1}) =(11, 491)$, and has the largest value of $h^{2,
  1}$ possible for any EFS CY threefold.

There are few ways available
to tune  an enhanced gauge group over the base $B_2 = \F_{12}$.
The gauge algebra on the curve $S$ with $S\cdot S= -12$ is $\ge_8$ and
cannot be enhanced.  Tuning a gauge algebra on any fiber $F$ would
increase the degree of vanishing at the point $S\cdot F$ beyond $(4,
5, 10)$, which is not allowed since such a point lies on $S$ and cannot
be blown up to give a valid base.  The only option for tuning is on
the curve $\tilde{S} = S+ 12F$, which has self-intersection $+ 12$ (or on
curves with a multiple of this divisor class, which would have
self-intersection $\geq 48$).  Tuning an $\gsu(2)$ factor on the
curve $\tilde{S}$ gives 88 fundamental matter fields, from
Table~\ref{t:matter-1}, so the Hodge numbers are
$(12, 318)$.  A threefold with these Hodge numbers is in the Kreuzer-Skarke
database, but has $h^{2, 1}< 350$, so we do not concern ourselves
  further with it here.  Tuning any larger gauge group factor reduces $h^{2, 1}$
still further; for example, tuning an $\gsu(3)$ gives Hodge numbers
$(13, 229)$.

This example illustrates the basic paradigm: on curves of higher
self-intersection, there are fewer restrictions on the possible
tunings, but more charged matter is required to fulfill anomaly
cancelation conditions. As a rule of thumb, it is often easy to
increase $h^{1, 1}$ via tuning so long as one is willing to accept a
large decrease in $h^{2, 1}$.

There is one other possibility that should be discussed here, and that
is the possibility of tuning an abelian gauge group factor.  As shown
in \cite{mt-sections}, any $U(1)$ factor 
can be seen as arising from a
Higgsed $SU(2)$ gauge group factor 
(which may be a subgroup of a larger nonabelian group),
under which some matter transforms
in the adjoint representation.  The $U(1)$ factor is associated with
the divisor class $C$ in the base that supports the $SU(2)$ gauge
group after unHiggsing; to have an adjoint, irreducible curves in this
divisor class must have nonzero genus.  In the case of $B_2 =
\F_{12}$, the divisor class $C$ cannot intersect $S$ without producing
a $(4, 6)$ singularity, so it must be a multiple $C = n \tilde{S}$ of
the curve of self-intersection $+ 12$ in $B_2$.  For $n = 2$, the
curve $2 \tilde{S}$ has genus $g = 11$, and the resulting $SU(2)$
model would have $11$ adjoint matter fields and 128 fundamental matter
fields.  Although this model should exist, it has a substantially
reduced number of Weierstrass moduli corresponding to uncharged matter
fields, even after breaking of the $SU(2)$ by a single adjoint.
Similarly, a discrete abelian group would involve further breaking of
the $U(1)$ that would maintain a relatively small value of $h^{2, 1}$.
Thus, while in principle it may be possible to tune an abelian factor,
for this base and the others considered here the resulting Calabi-Yau
threefold has relatively small $h^{2, 1}$, and we do not need to
consider abelian factors in constructing threefolds with $h^{2, 1}
\geq 350$.  We discuss abelian factors further in
\S\ref{sec:issues-abelian}.

\subsection{Tuning models over $\F_{8}$ and $\F_{7}$}
\label{sec:tuning-78}

The generic elliptically fibered Calabi-Yau threefolds over the
Hirzebruch bases $\F_7$ and $\F_8$ have Hodge numbers $(10, 376)$.
The discussion of tuning over these bases is precisely analogous to
the preceding discussion for the base $\F_{12}$, and there are no
tuned models over these bases with $h^{2, 1}\geq 350$.  Since $376-29
< 350$, there are also no threefolds formed over bases that are
  blow-ups of $\F_7$ or $\F_8$ that have $h^{2, 1}\geq 350$.  The
  threefolds with Hodge numbers $(10, 376)$ over these bases are,
  however, continuously connected to the threefolds over $\F_{12}$ and
  blow-ups thereof; for example, blowing up $\F_8$ at four generic
  points on the curve $S$ of self-intersection $-8$ gives a base that
  is equivalent to the one reached by blowing up $\F_{12}$ at four
  generic points.  It is not immediately clear whether the threefolds
  formed from generic elliptic fibrations over $\F_7$ and $\F_8$ are
  equivalent.  We discuss this issue further in \S\ref{sec:equivalence}.

\subsection{Decomposition into fibers}

To find further EFS CY threefolds with large $h^{2, 1}$ we must blow up
one or more points in the base $B =\F_{12}$ to get further bases over
which a variety of Weierstrass models can be tuned.  We can blow up
any point on $\F_{12}$ that does not lie on the curve $S$ of
self-intersection $-12$.  Any such point lies on a fiber $F$ that
intersects $S$ and $\tilde{S}$ each at one point.  After blowing up one point
we can blow up another point on the same fiber or on another fiber.
Until the number of blow-ups is large ($> 12$), blow-ups on distinct
fibers do not interact, so that we may analyze the sequence of
blow-ups possible along one given fiber, and then we can combine such
sequences to construct threefolds involving the blow-ups of multiple
fibers.  Along any given fiber, as long as each blow-up occurs at an
intersection of curves of negative self-intersection or at the point
of intersection of the fiber with $\tilde{S}$, we can use toric
methods for describing the monomials, as in 
\S\ref{sec:Weierstrass}.
After a sufficient number of blow-ups, it is also possible to
construct fibers that do not fit into the toric framework, though we
need to consider only one example of this in the analysis for
threefolds with $h^{2, 1}\geq 350$.
For fibers that simply consist of a linear sequence of mutually
intersecting curves, such as those in Figure~\ref{f:blow-up}, for
convenience we label the curves $C_1, C_2, \ldots$, with $C_1$ the
curve that intersects the $-12$ curve $C$ (so we always have $C_1
\cdot C_1 = -1$).

\subsection{$\F_{12}$ blown up at one point $(\F_{12}^{[1]})$}
\label{sec:12-1}

We now consider the sequence of fiber geometries that can arise when
we blow up consecutive points in $\F_{12}$ that lie in a single fiber.
Blowing up a generic point on $\F_{12}$ gives a toric base with a
single nontrivial fiber $(-1, -1)$ containing curves $C_2, C_1$, as in
the first step in Figure~\ref{f:blow-up}.  As discussed in
\S\ref{sec:bases}, blowing up a point when no gauge groups are
involved leads to a shift in Hodge numbers of $+1, -29$.  The generic
elliptic fibration over the base $\F_{12}$ with a single blow-up,
which we denote $\F_{12}^{[1]}$, thus has Hodge numbers $(12, 462)$.

For the base $\F_{12}^{[1]}$, as for $\F_{12}$, there is no place that
we can tune a gauge group other than the $+ 11$ curve; as described in
\S\ref{sec:Weierstrass}, tuning an $\gsu(2)$ factor on either $-1$
curve raises the degree of vanishing of $f, g$ on $S$ to $(4, 6)$, and
is not possible.  Any other tuning on the $-1$ curves increases the
degree of vanishing still further and is not allowed.  The model with
an $\gsu(2)$ on the $+ 11$ curve is just the blow-up of the case with
Hodge numbers $(12, 318)$ and has Hodge numbers $(13, 301)$
(note that the number of fundamental matter fields is reduced by 6
compared to the $+ 12$ curve in $\F_m$).

\subsection{Threefolds over the base $\F_{12}^{[2]}$}

Now consider blowing up a second point on $\F_{12}$ by blowing up a
point on $\F_{12}^{[1]}$.  If the second point is a generic point that
does not lie on the first blown-up fiber, we can take it to be on a
separate fiber.  The shift in Hodge numbers just adds between the two
fibers and is then $2 \times (+1, -29)$, giving an EFS threefold with
Hodge numbers $(13, 433)$.

Now, consider which points in the $(-1, -1)$ fiber can be blown up and
give a consistent model.  We cannot blow up a point in $C_1$ (the $-1$
curve intersecting the $-12$ curve), since then it would become a $-2$
intersecting a $-12$, which is not allowed by the intersection rules
of \cite{clusters}.  A representative $\tilde{S}'$ of the (non-rigid)
$+ 11$ class passes through each point on $C_2$ (this is one of the
degrees of freedom in $w_{\rm aut}$ in (\ref{eq:monomials-w})), so
without loss of generality we can blow up any point in $C_2$, and we
get a fiber of the form $(-1, -2, -1)$, which now connects a $+ 10$
curve $\tilde{S}'$ to a $-12$ curve.  In the absence of tuning, the
corresponding Calabi-Yau threefold simply lies in a codimension one
locus in the moduli space of complex structures of the threefold with
Hodge numbers $(13, 433)$ having two $(-1, -1)$ fibers.  Now, however,
we consider what can be tuned on the $(-1, -2, -1)$ fiber.  For the
same reason, described in the example in \S\ref{sec:Weierstrass}, that
we could not tune any gauge group on the upper $-1$ of a $(-1, -1)$
fiber, we cannot tune a gauge group on the $-2$ curve $(C_2)$.  Thus,
the only curve on which we can tune any gauge group is the top $-1$
curve $C_3$.  It is easy to check that we can tune an $\gsu(2)$ on this
top curve, either by tuning a type $I_2$ singularity or the more
specialized type $III$.  This does not violate the Zariski or anomaly
conditions, and explicit examination of the Weierstrass model shows
that this configuration is allowed.  From Table~\ref{t:matter-1}, we
see that this tuning shifts the Hodge numbers by $(+ 1, -17)$, giving
a Calabi-Yau with Hodge numbers $(14, 416)$.  No other Calabi-Yau with
$h^{2, 1}\geq 350$ can be formed by tuning a gauge group over
$\F_{12}^{[2]}$.  Some checking is needed, however, to confirm that no
other gauge group can be tuned on $C_3$.  From the analysis of
\S\ref{sec:Weierstrass}, the only allowed degrees of vanishing on
$C_2$ are $(0, 0, 1)$ or $(1, 1, 2)$, so by the averaging rule the
only way in which the degrees of vanishing on $C_3$ could give any
larger gauge algebra than $\gsu(2)$ is for a type $IV$ $(2, 2, 4)$
singularity carrying an $\gsu(3)$ gauge algebra.  Expanding $g = g_0 (w) +
g_1 (w) \zeta + \cdots$ in powers of a coordinate $\zeta$ that
vanishes on $C_3$ (with $w = 0$ on $\tilde{S}$), the
condition for an $\gsu(3)$ gauge group at a type $IV$ singularity
is that $g_2$ be a perfect
square.  The highest power of $w$ appearing in $g_2$, however, is
$w^7$, corresponding to the single monomial of degree $5$ in $g$ over
$S$.  If $g_2$ is a perfect square then this coefficient would have to
vanish, giving a $(4, 6)$ vanishing on $S$.  Thus, there is only one
possible tuning of $\F_{12}^{[2]}$, with a single $\gsu(2)$ on $C_3$.

\subsection{Threefolds over the base $\F_{12}^{[3]}$}

Now we consider blowing up a third point on $\F_{12}$.  Unless all
three points are on the same fiber, we simply have a combination of
the previously considered configurations.  On the twice blown up fiber
$(-1, -2, -1)$, we cannot blow up on $C_1$ or $C_2$, or we would have
a cluster that is not allowed in such close proximity to the $-12$
through the rules of \cite{clusters}.  So we can only blow up on the
initial $-1$ curve $C_3$.  As above, a representative of the $+ 10$
curve on $\F_{12}^{[2]}$ passes through each point on $C_3$, so a
blow-up at any such point gives the base $\F_{12}^{[3]}$ with fiber
$(-1, -2, -2, -1)$.  This is on the same moduli space as the
Calabi-Yau with Hodge numbers $(14, 404)$ having three $(-1, -1)$
fibers.  We can, however, tune various gauge groups on $\F_{12}^{[3]}$
that fix the $-2$ structure in place.  From the analysis of previous
cases we know that we cannot tune a gauge group on $C_1$ or $C_2$, and
the only possible gauge  algebra on $C_3$ is $\gsu(2)$.  (Note that the
argument from the previous section constraining the gauge group on
$C_3$ remains valid even when additional points are blown up).  By the
averaging rule, the largest possible vanishing orders of $f, g,
\Delta$ that are possible on $C_4$ are $(3, 3, 6)$.  A systematic
analysis shows that we can tune the following gauge algebra
combinations on the initial $(-1, -2)$ curves $C_4$ and $C_3$:
\begin{eqnarray}
\cdot \oplus\gsu (2) &\rightarrow &  (h^{1, 1},h^{2, 1}) = (15, 399)
\label{eq:3-02}\\
\gsu (2) \oplus  \cdot &\rightarrow &  (h^{1, 1},h^{2, 1}) = (15, 387)
\label{eq:3-20}\\
\gsu (2)\oplus\gsu (2) &\rightarrow &  (h^{1, 1},h^{2, 1}) = (16, 386)
\label{eq:3-22}\\
\gsu (3)\oplus\gsu (2) &\rightarrow &  (h^{1, 1},h^{2, 1}) = (17, 377)
\label{eq:3-32}\\
\gg_2 \oplus\gsu (2) &\rightarrow &  (h^{1, 1},h^{2, 1}) = (17, 371)
\label{eq:3-g2}
\end{eqnarray}
Note that in the last three cases, there is bifundamental matter.
For example, in the case (\ref{eq:3-22}) the shift in $h^{2, 1}$
corresponds to the  net change in
$V-H_{\rm ch}$.  From Table~\ref{t:matter-1} we would expect
$-5-17 = -22$, but there is a bifundamental ${\bf 2} \times {\bf 2}$
from the intersection 
between the $-1$ and $-2$ curves so that 4 of the matter
hypermultiplets have been counted twice,  and the actual change to
$h^{2, 1}$ is $404-18 = 386$.  

All of the tunings (\ref{eq:3-02})-(\ref{eq:3-g2}) give consistent
constructions of EFS Calabi-Yau threefolds.  Note, however, that
the threefold realized through (\ref{eq:3-20}) is not a generic
threefold in the given branch of the moduli space.  For this
construction, the curve $C_2$ is a $-2$ curve without vanishing degree
for $\Delta$.  Thus, the threefold can be deformed by moving $C_1$ to
a different point on $S$.  This gives a $\C^*$ base with a single
$(-1, -1)$ fiber and  a $(-1, -2, -1)$ fiber with a single $\gsu (2)$
as can be tuned on $\F_{12}^{[2]}$.  Checking the Hodge numbers, we
see that indeed the resulting model is equivalent to the blow-up of
the $(14, 416)$ threefold at a generic point, so we do not list this
construction separately in Table~\ref{t:Hodge-350}.

The final case (\ref{eq:3-g2}) is of particular interest, as it
appears to give a Calabi-Yau threefold that did not arise in the
complete classification by Kreuzer and Skarke of threefolds based on
hypersurfaces in toric varieties.  In this case there is a matter
field charged under the $\gg_2 \oplus\gsu (2)$ transforming in the
$({\bf 7},\frac{1}{2} {\bf 2})$ (half-hypermultiplet in the
fundamental of $\gsu (2)$), which raises $h^{2, 1}$ by 7: $404-5- 35 +
7 = 371$.  Given the apparent novelty of this construction, for this
particular threefold we spell out some of the details of the
Weierstrass monomial calculation that we have performed as a
cross-check.  After requiring that $(f,g,\Delta)$ vanish to degree
$(2,3,6)$ on the $(-1)$-curve $C_4$ and $(2,2,4)$ on the adjacent
$(-2)$-curve $C_3$ ($\Delta$ must vanish to degree 4 on $C_3$ and to
degree $2$ on $C_2$, by the averaging rule), the number of Weierstrass
monomials in $f, g$ becomes
\begin{equation}
W_f=125,\ \ \ W_g=260 \;.%W_{extra}=((9+1)+(0+1)+2)=13,
\end{equation}
With $w_{\rm aut} = 1 + (9 + 1) = 11, N_{-2} = G_1 = 0$, we have then
$h^{2, 1}=125+260-11-3=371$, in agreement with the expectation from
anomaly cancellation. 
It is also straightforward to check that this set of Weierstrass
monomials does not impose any unexpected $(4, 6)$ vanishing on curves
or points in the base which would invalidate the threefold construction.
Because a
$(2, 3, 6)$ tuning is ambiguous,   we consider the possible
monodromies associated with the gauge group on $C_4$, which can be
analyzed in terms of monomials in a local coordinate system.
Expanding $f=\sum_i\hat{f}_i\zeta^i$ and $g=\sum_i \hat{g}_i\zeta^i$ in a coordinate $\zeta$
that vanishes on $C_4$, the monodromy that determines the choice of
gauge algebra $\gg_2,\gso (7)$ or $\gso (8)$ is determined by the
form the polynomial containing the leading
order terms in $\zeta$ from
the Weierstrass equation
\begin{equation}
x^3+\hat{f}_2x+\hat{g}_3 \,,
\label{eq:w-23}
\end{equation}
where the coefficients $\hat{f}_2$ and $\hat{g}_3$ are functions 
 on the $-1$ curve $C_4$ only of the usual
coordinate $w$,
which vanishes on $\tilde{S}$.  The monodromy condition that
selects the gauge group can be found from the factorization structure
of \eq{eq:w-23},
\begin{eqnarray}
x^3+Ax+B  \text{\ \ (generic)} & \Rightarrow & \gg_2 \nonumber \\
(x-A)(x^2+Ax+B) & \Rightarrow & \gso(7) \label{eq:factorize-7} \\
(x-A)(x-B)(x+(A+B)) & \Rightarrow & \gso(8) \label{eq:factorize-8} \,.
\end{eqnarray}
From an analysis similar to that described
in section \S\ref{sec:Weierstrass} (which can also be read off directly
from Figure~\ref{f:example-monomials}, noting that the monomials 
$\zeta^j w^k$ correspond to $z^{j+ 3 (n-k)}w^k$, for $n = 4, 6$ for
$f, g$ respectively), we find that $\hat{f}_2, \hat{g}_3$ have the form
\begin{eqnarray}
\hat{f}_2(w) & = & \hat{f}_{2,0}+\hat{f}_{2,1}w+\hat{f}_{2,2}w^2+\hat{f}_{2,3}w^3+\hat{f}_{2,4}w^4 \nonumber \\
\hat{g}_3(w) & = &
\hat{g}_{3,0}+\hat{g}_{3,1}w+\hat{g}_{3,2}w^2+\hat{g}_{3,3}w^3+\hat{g}_{3,4}w^4+\hat{g}_{3,5}w^5+
\hat{g}_{3,6}w^6+\hat{g}_{3,7}w^7\,.
\end{eqnarray}
The $w^7$ term in $g_3$ corresponds to the monomial $w^7$ with
coefficient $g_{0, 7}$ in the original $z, w$ coordinates, which as
discussed above cannot be tuned to zero since this would force a
$(4, 6, 12)$ singularity on $S$.  This implies,
however, that (\ref{eq:w-23}) cannot have a nontrivial factorization.
Any tuning of an $\gso(7)$ gauge algebra, for example, must, upon
expanding (\ref{eq:factorize-7}), yield $\hat{f}_2=B-A^2$, which would
imply that $A$ must be no more than quadratic and $B$ no more than
quartic (a higher-order cancellation with $A$ cubic and $B$ sextic is
not possible since this would lead to 9th order terms in $g$).  This
means, however, that $\hat{g}_3=AB$ could be at most of order six; in
other words, this factorization cannot be achieved without tuning the
$w^7$ term in $\hat{g}_3$ to zero. A similar argument demonstrates
that an $\gso(8)$ cannot be tuned on $C_4$, but it is clear already
that any tuning of $\gso(8)$ involves at least the restrictions of
$\gso(7)$ on the monomials in question, hence the impossibility of
$\gso(7)$ implies the impossibility of $\gso(8)$.  Thus, the presence
of the $w^7$ term in $g$ guarantees that the monodromy associated with
a Kodaira type $I_0^*$ singularity over $C_4$ must give an $\gg_2$
gauge algebra, as in (\ref{eq:3-g2}).  

The upshot of this analysis is that the tuning (\ref{eq:3-g2}) seems
to give a threefold with Hodge numbers $(17, 371)$ while no tuning
beyond $\gg_2$ is possible on $C_4$.  Some possible subtleties in the
interpretation of the $(17, 371)$ threefold are discussed in
\S\ref{sec:new}.

One other issue that should also be explained explicitly is the reason
that it is not possible to tune an $\gsu (3)$ algebra on $C_4$ without
tuning a gauge group on $C_3$.  It is straightforward to check using
monomials that tuning $g$ to vanish to order 2 on $C_4$ forces $g$ to
also vanish to order $2$ on $C_3$, so a type $IV$ (2, 2, 4) vanishing
on $C_4$ forces a type $III$ (1, 2, 3) vanishing at least on $C_3$,
which must always be associated with a nonabelian gauge group.  And
tuning a $(0, 0, 3)$ vanishing on $C_4$ produces at least a $(0, 0,
2)$ vanishing on $C_3$ by the averaging rule, but by checking
monomials we can verify that no vanishing is imposed on $f$ or $g$ on
$C_3$ so again tuning an $\gsu (3)$ on $C_4$ necessarily imposes at
least an $\gsu (2)$ on $C_3$.

This completes the
classification of possible tuning structures that are possible on
$\F_{12}^{[3]}$; the resulting Calabi-Yau threefolds are tabulated in
Table~\ref{t:Hodge-350}.

\subsection{Threefolds over the base $\F_{12}^{[4]}$}

At the next stage, again we can only blow up on the first $-1$ 
curve ($C_4$)
in the $(-1, -2, -2, -1)$ fiber in $\F_{12}^{[3]}$, since for example
a $-12$ curve cannot be connected by a $-1$ curve to a $(-2, -3)$
cluster so we cannot blow up on the second ($-2$) curve $C_3$.  Again,
the Calabi-Yau threefold $\F_{12}^{[4]}$ over the base with the
resulting $( -1, -2, -2, -2, -1)$ fiber is in the same moduli space as
the one with four $(-1, -1)$ fibers and has Hodge numbers (15, 375).
But there are an increasing number of possible gauge groups that can
be tuned on the initial three curves $C_5, C_4,$ and $C_3$.

All of the analysis performed for tunings on $C_4$--$C_1$ in
$\F_{12}^{[3]}$ holds for tunings on these same curves in
$\F_{12}^{[4]}$.  Thus, each of the gauge groups tuned over
$\F_{12}^{[3]}$ can be tuned in a parallel fashion on $\F_{12}^{[4]}$.
The only difference is that $C_4$ is now a $-2$ curve, so the gauge
groups on that curve have reduced matter content and the change in
Hodge numbers from the tuning decreases accordingly.  For example,
while tuning an $\gsu(2)\oplus \gsu(2)$ on $C_4$ and $C_3$ over
$\F_{12}^{[3]}$ shifts the Hodge numbers by $(\Delta h^{1,1}, \Delta
h^{2,1})=(+2,-18)$ as discussed above, the shift for the same gauge
group tuning on $\F_{12}^{[4]}$ is $(+ 2, -6)$ since there are 6 fewer
matter fields in the fundamental ${\bf 2}$ representation of the $\gsu
(2)$ over $C_4$.  We can also confirm directly that none of the
allowed tunings on $C_4$--$C_1$ impose any mandatory vanishing
condition on $C_5$.  Thus, the tunings
(\ref{eq:3-22})--(\ref{eq:3-g2}) can all be done in a similar fashion,
giving another set of threefolds tabulated in Table~\ref{t:Hodge-350},
including another apparently new threefold not in the CY database at
$(18, 363)$.  Note that the tuning (\ref{eq:3-02}) of a single $\gsu
(2)$ on $C_3$ in $\F_{12}^{[4]}$ gives a threefold on the same moduli
space as the blow-up at a generic point of this tuning on
$\F_{12}^{[3]}$, with Hodge numbers $(16, 370)$.

Finally, we can consider tuning a gauge group on $C_5$ in combination
with any other gauge groups on the other curves.  As in the analysis
in the previous section, if no gauge group is tuned on $C_3$, the
threefold is non-generic since the curve $C_1$ can be moved on $S$.
By the averaging rule, tuning an $\gsu(2)$ on both $C_3$ and $C_5$
will also force an $\gsu(2)$ on $C_4$.  An $\gsu(2)$ can be tuned on
$C_5$, along with $\gsu(2)$ factors on $C_4$, and $C_3$, giving a
threefold with Hodge numbers $(18, 356)$.  Enhancement of
the $\gsu(2)$ on $C_4$ to $\gsu(3)$ is then still possible, which
yields a threefold with the
Hodge numbers $(19,355)$.  Note that these Hodge numbers are
identical to those of a generic fibration over a five-times blown up
$\F_{12}$ (discussed below); this provides the first example of a
situation where two apparently distinct constructions produce
threefolds with identical Hodge numbers.  The possible
relationship between such models
is discussed in \S\ref{sec:equivalence}. Finally, the middle $\gsu(3)$ (on
$C_4$) can again be enhanced to $\gg_2$, yielding a model with Hodge numbers
$(19,353)$,  another new construction that does not appear in the
Kreuzer-Skarke database. 
There are also a number
of possible configurations where $\gsu (3)$ and larger gauge groups
are tuned on $C_5$, but since $C_5$ is a $-1$ curve and gauge groups tuned on such divisors carry more matter, these all give threefolds with smaller Hodge
numbers $h^{2, 1}< 350$.  One such tuning that is worth mentioning,
however, is given by imposing the condition that $\Delta$ vanish to degree 4 on $C_5$.
This can be arranged, giving for example a model with gauge group
$\gsp(2)\oplus \gsu(2)$ and Hodge numbers $(20,340)$, which arises in
the Kreuzer-Skarke database.
A more detailed exploration of
these and other models with $h^{2, 1}< 350$ is left to further work.
  This completes the summary of threefolds based on tuning of
$\F_{12}^{[4]}$.

\subsection{Five blow-ups}

At this stage the story becomes  even more interesting.  We can blow up
the fiber $(-1, -2, -2, -2, -1)$
again at an arbitrary point on $C_5$ to get (A) $\F_{12}^{[5]}$ with a
resulting $(-1, -2, -2, -2, -2, -1)$ fiber.  We can also, however,
blow up in two other ways.  We can blow up the point of intersection
between $C_5$ and $C_4$
giving a chain (B) $(-2, -1, -3, -2, -2, -1)$.  Alternatively, we can
blow up a generic point in the curve $C_4$, giving the fiber
(C) shown in Figure~\ref{f:non-toric-fiber}.  In the latter case, the
fiber and associated base no longer have a toric description.  Let us
consider these three cases in turn:

\begin{figure}
\begin{center}
\includegraphics[width=9cm]{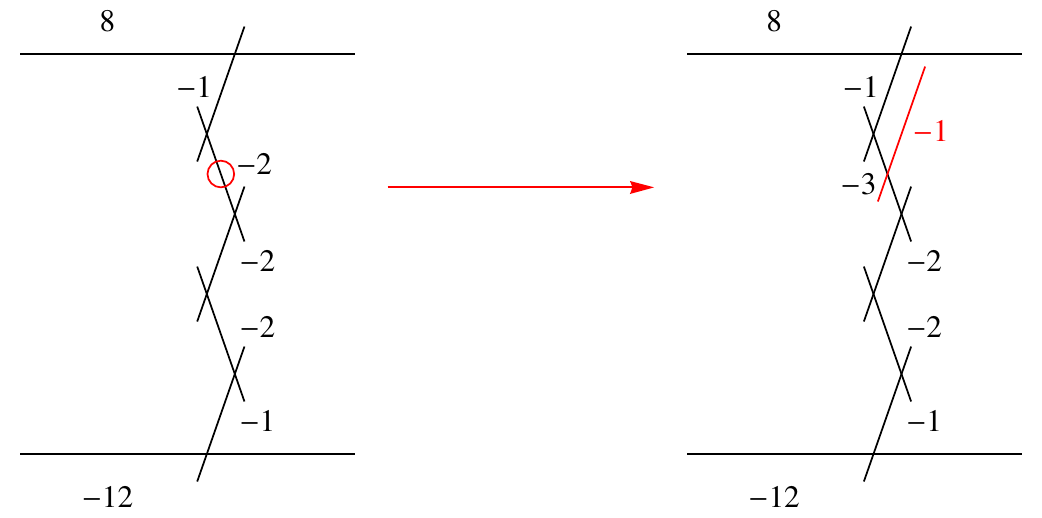}
\end{center}
\caption[x]{\footnotesize  Blowing up on the top $-2$ curve ($C_4$) on
a $(-1, -2, -2, -2, -1)$ chain results in a divisor structure giving a
non-toric base, with a $(-3, -2, -2)$ non-Higgsable cluster
(case (C) in the text).
In the limit of moduli space where
the intersection points of the two $-1$ curves with the $-3$ curve
coincide, the fiber becomes $(-2, -1, -3, -2, -2, -1)$ and the base
becomes toric (case (B)).}
\label{f:non-toric-fiber}
\end{figure}

(A) This is the straightforward generalization of the previous
examples, $\F_{12}^{[5]}$ has Hodge numbers $(16, 346)$, and is on the
same moduli space as the base with five $(-1, -1)$ fibers.  There are
a variety of tunings, which all have $h^{2, 1}< 350$ and which
  therefore for the present purposes we omit.  It bears mentioning
  that tunings on the multiple $-2$ curves in this base give a rich
  variety of possible threefolds, and it is at this point that larger
algebras such as $\gf_4$ and $\ge_6$ can be tuned.

(B) In this case, as discussed in \cite{toric}, the appearance of
the non-Higgsable cluster $(-3, -2)$ requires a non-Higgsable gauge
algebra $\gg_2 \oplus\gsu (2)$.  The associated rank 3 gauge algebra
with 17 vector multiplets and 8 charged matter multiplets raise the
Hodge numbers of this base to $(19, 355)$. 
There are various tunings on $C_6$, but all go below $h^{2, 1}= 350$,
except for a single $\gsu(2)$ on $C_6$ that gives a standard shift to
a threefold with Hodge numbers
$(19, 355) + (1,-5)=(20,350)$. 
Note that without tuning, the initial $-2$ curve in this fiber
represents an extra Weierstrass modulus, so this is not a generic
configuration, as discussed in the following case.
Note also, however, that the analysis of section \S\ref{sec:tuning}
shows that no gauge group can be tuned on the $-1$ curve $C_5$ since
it is adjacent to a non-Higgsable cluster that is not a single $-3$ curve.

(C) In this non-toric case we again have the same non-Higgsable cluster as in
the previous case, and the same Hodge numbers $(19, 355)$.  In this
case there are also no tunings possible.  
This construction represents the generic class of threefolds of which
the untuned model (B) above represents a codimension one limit.  The
final blow-up of a point in $C_4$ in this non-toric construction can
be taken to approach the point which was blown up to form $C_5$ in
$\F_{12}^{[4]}$, producing the $-2$ curve found in (B).

\subsection{More blow-ups}

Further blow-ups raise the Hodge number $h^{2, 1}$ below 350.  As the
number of blow-ups increases, the number of fiber configurations also
increases.  We leave a systematic analysis of tuned models over
further blown up bases for further work.

\begin{table}
\begin{center}
\begin{tabular}{| cc | c | c | c | c | c  | }
\hline
$h^{11}$ & $h^{21}$ & Base & K-S $\#$ & $(\Delta h^{11},\ \Delta h^{21})$ &
 Fiber & ${\cal G}_{\rm extra}$ \\
\hline
11 & 491 & $\F_{12}$ & 1 & (0,0) & 0 &  \\
12 & 462 & $\F_{12}$ & 2 & $(1,-29)$& 11 &  \\
13 & 433 & $\F_{12}$ & 4 & 2 $\times$ $(1,-29)$ &$2 \times $(11)$$ &  \\
14 & 416 & $\F_{12}$ & 2 & $(3,-75)$ & $\underbar{1}21$ & $\gsu(2)$  \\
14 & 404 & $\F_{12}$ & 6 & 3 $\times$ $(1,-29)$ & $3 \times $(11)$$ &  \\
15 & 399 & $\F_{12}$ & 1 & $(4,-92)$ & $1\underbar{2}21$ & $\gsu(2)$ \\
15 & 387 & $\F_{12}$ & 4 & (1, -29) $+$ (3, -75) & 11 +
$\underbar{1}21$ & $\gsu(2)$ \\
16 & 386 & $\F_{12}$ & 1 & $(5,-105)$ & \underbar{1}$\underbar{2}21$ & $\gsu(2)\oplus \gsu(2)$ \\
17 & 377 & $\F_{12}$ & 3 & $(6,-114)$ & \underbar{1}$\underbar{2}21$ & $\gsu$(3)$\oplus \gsu(2)$ \\
10 & 376 & $\F_8$ & 2 & $(0,0)$ & 0 & \\
& & $\F_7$ &  & $(0,0)$ & 0 & \\
15 & 375 & $\F_{12}$ & 9 & 4 $\times$ $(1,-29)$ & $4 \times $(11)$$ &  \\
17 & 371 & $\F_{12}$ & 0  & $(6,-120)$ & \underbar{12}21 & $\gg_2\oplus \gsu(2)$ \\
16 & 370 & $\F_{12}$ & 3 & $(1,-29)$ $+$ $(4,-92)$ & 11 + $1\underbar{2}21$ & $\gsu $(2)$$ \\
17 & 369 & $\F_{12}$ & 1 & $(6,-122)$ & 1\underbar{22}21 & $\gsu(2)\oplus \gsu(2)$ \\
18 & 366 & $\F_{12}$ & 2 & $(7,-125)$ & 1\underbar{22}21 & $\gsu(3)\oplus \gsu(2)$ \\
18 & 363 & $\F_{12}$ & 0 & $(7,-128)$ & 1\underbar{22}21 & $\gg_2 \oplus \gsu(2)$ \\
16 & 358 & $\F_{12}$ & 7 & 2  $\times$ (1, -29) $+$
$(3, -75)$ & $2 \times (11) +$\underbar{1}21 & $\gsu(2)$ \\
17 & 357 & $\F_{12}$ & 2 & $(1, -29) + (5, -105)$ & $11 +$
\underbar{12}21 & $\gsu(2)\oplus \gsu(2)$ \\
18 & 356 & $\F_{12}$ & 1 & $(7,-135)$ & \underbar{122}21 & $\gsu(2)\oplus \gsu(2)\oplus \gsu(2)$ \\
19 & 355 & $\F_{12}$ & 3 & $(8,-136)$ & $21\overline{32}21$ &
$\gg_2\oplus \gsu(2)$ \\
& & & & & & (generically non-toric)\\
& & $\F_{12}$ & 3 & $(8,-136)$ & \underline{122}21 & $\gsu(2)\oplus\gsu(3)\oplus\gsu(2)$ \\
19 & 353 & $\F_{12}$ & 0 & $(8,-138)$ & $\underbar{122}21$ & $\gsu(2)\oplus\gg_2\oplus \gsu(2)$\\
20 & 350 & $\F_{12}$ & 1 & $(9,-141)$ & $\underbar{2}1\overline{32}21$ & $\gsu(2)\oplus \gg_2\oplus \gsu(2)$ \\
\hline
\end{tabular}
\end{center}
\caption[x]{\footnotesize Table of all possible
  Calabi-Yau threefolds that are elliptically fibered with section and
  have $h^{21} \geq 350$. For each pair of Hodge numbers, the number
  of distinct constructions found by Kreuzer and Skarke giving these
  Hodge numbers is listed (0= new construction).  The data for
  explicit construction through a tuned elliptic fibration over a
  blow-up of $\F_{12}$ is given for each threefold.  
In each case, the fiber types and extra tuned gauge groups (beyond
those forced from the structure of the original Hirzebruch base --
$\ge_8$ in all cases except the $(10, 376)$ CY's) is indicated.
Each fiber is given by a sequence of the (negative of the)
self-intersection numbers of the curves in the fiber; underlined
curves carry tuned gauge group factors, while overlined  curves carry
gauge group factors associated with non-Higgsable clusters.}
\label{t:Hodge-350}
\end{table}

\section{Conclusions}
\label{sec:conclusions}

In this paper we have initiated a systematic analysis of the set of
all elliptically fibered Calabi-Yau threefolds, starting with those
having large Hodge number $h^{2, 1}$.  These Calabi-Yau threefolds fit
together into a single connected space, with the continuous moduli
spaces associated with different topologies connected together through
transitions between singular points in the different components of the
moduli space.  This structure is clearly and explicitly described in
the framework of Weierstrass models.  In principle, the approach  taken
here could be used to classify all EFS CY threefolds.  There are,
however, a number of practical and technical limitations to carrying
out this analysis for the set of all threefolds with arbitrary Hodge
numbers given the current state of knowledge.  We describe these
issues in \S\ref{sec:limitations-3}.  A similar analysis could in
principle be carried out for Calabi-Yau fourfolds, though in this
context there are even larger unresolved mathematical questions,
discussed in \S\ref{sec:limitations-4}.  Some other comments on future
directions are given in \S\ref{sec:future}.

\subsection{Classifying all EFS Calabi-Yau threefolds}
\label{sec:limitations-3}

In order to classify the complete set of EFS Calabi-Yau threefolds,
some specific technical problems that begin to arise at smaller Hodge
numbers need to be resolved.  The primary outstanding issues seem to
be the following 4 items:
\vspace*{0.1in}

\noindent
{\bf General bases:}
A systematic means for  explicitly enumerating the complete set of
possible bases $B_2$, including bases that are neither toric nor
``semi-toric'' has not yet been developed.
\vspace*{0.1in}

\noindent
{\bf Tuning classical groups:}
A general rule for determining when the gauge groups $SU(N)$, $SO(N),$
and $Sp(N)$ can be tuned on a given divisor is not known.
\vspace*{0.1in}

\noindent
{\bf Codimension two singularities:}
A complete classification of codimension two singularities and
associated matter representations has not yet been realized.
\vspace*{0.1in}

\noindent
{\bf Extra sections and abelian gauge group factors:}
There is no general approach available yet for determining when an
elliptic fibration of arbitrary Mordell-Weil rank can be tuned over a
given base $B_2$.
\vspace*{0.1in}

We describe these issues in some further detail and summarize the
current state of understanding for each issue in the remainder of this
section.  If all these issues can be resolved, it seems
that the complete classification and enumeration of EFS CY threefolds
may be a problem of tractable computational complexity, as discussed
further in \S\ref{sec:issues-base}.

\subsubsection{General bases}
\label{sec:issues-base}

As discussed in \S\ref{sec:bases}, the set of possible bases is
constrained by the set of allowed non-Higgsable clusters of
intersecting divisors with negative self-intersection \cite{clusters},
and a complete enumeration of all bases with toric and semi-toric
($\C^*$) structure has been completed \cite{toric, Martini-WT}.  In
principle, there is no conceptual obstruction to explicitly
enumerating the finite set of possible bases $B_2$ that support an
elliptically fibered CY threefold, but in practice this becomes rather
difficult since the intersection structure can become rather
complicated as more points are blown up.  For bases with smaller
values of $h^{2, 1}$ than those considered here, there are more ways
in which points can be blown up without preserving a toric or $\C^*$
structure.  This leads to increasingly complicated branching
structures in the set of intersecting divisors.  It is a difficult
combinatorial problem to track the new divisors of negative self
intersection that may appear as non-generic points are blown up in a
base that has no $\C^*$ symmetry.  For example, new curves of negative
self intersection may appear from curves of positive or vanishing self
intersection that pass through multiple blown up points; in more
extreme cases, a set of points may be blown up that lie on a highly
singular codimension one curve, complicating the divisor intersection
structure.  A related issue is that the number of generators of the
Mori cone of effective divisors can become large -- indeed, for the
del Pezzo surface $dP_9$ formed by blowing up $P^2$ at 9 generic
points, the cone of effective divisors is generated by an infinite
family of distinct $-1$ curves.

While the combinatorics of this problem may seem forbidding, several
pieces of evidence suggest that a complete enumeration may be a
tractable problem.  The analysis of $\C^*$ bases in \cite{Martini-WT}
shows that allowing certain kinds of branching and corresponding loops
in the web of effective rigid divisors (associated with multiple
fibers intersecting $S, \tilde{S}$) does not dramatically increase the
range of possible bases\footnote{some simple branching structures of
  this kind are also encountered in the classification of 6D
  superconformal field theories \cite{Heckman-mv}}; the full set of
$\C^*$ bases is several times larger than the number of toric bases
($\sim$ 160,000 vs.  $\sim$60,000), but not exponentially larger.  It
also seems that as the complexity of $\C^*$ bases increases, the range
and complexity of non-$\C^*$ structures that can be added by further
blow-ups decreases.  Further work in this direction is in progress,
but it seems likely that the total number of possible bases may not
exceed the number already identified as toric or $\C^*$-bases by more
than one or two orders of magnitude.

\subsubsection{Tuning $I_n$ and $I_n^*$ codimension one singularities}
\label{sec:issues-classical}

As described in \S\ref{sec:tuning}, \S\ref{sec:anomalies}, and
\S\ref{sec:Weierstrass}, though the intersection structure of
divisors, Zariski-type decomposition, and 6D anomalies can strongly
restrict which gauge groups can be tuned over any given configuration
of curves, these conditions are necessary but not sufficient, and to
verify that a valid threefold with given structure exists a more
direct method such as an explicit Weierstrass construction is needed.
For gauge algebras such as $\ge_7$ and $\ge_8$ that are realized by
tuning coefficients in $f$ and $g$ to get the desired Kodaira
singularity types, it is fairly straightforward to confirm that
Weierstrass models with the desired properties can be constructed.
For algebras like $\gg_2, \ge_6$, or $\gf_4$ that involve monodromy but
are still realized by tuning $f, g$, it is also possible to check the
Weierstrass model directly by considering the set of allowed
monomials in the specific model;  examples of this were
described in \S\ref{sec:systematic}.  There are some types of gauge
algebra, however, namely those realized by Kodaira type $I_n$ and
$I_n^*$, where the tuning required is on the discriminant $\Delta$ and
not directly on $f, g$.  This leads to a more difficult algebra
problem, since as $n$ becomes larger, the set of required conditions
become complicated polynomial conditions on the coefficients of $f,
g$, rather than linear conditions as arise in all other cases.

An example of this kind of difficulty arises in considering the tuning
of a Kodaira type $I_n$ singularity giving an $\gsu(n)$ gauge algebra
over a simple curve of degree one in the base $B_2=\P^2$.  In this
case, anomaly cancellation conditions restrict the rank of the group
so that $n \leq 24$.  But explicit construction of the models for large $n$
is algebraically somewhat complicated.  In this case, $f$ is a
polynomial of degree $12$ in local coordinates $z, w$ in the base, and
$g$ is a polynomial of degree 18.  If we consider a curve $C$ defined
by the locus where $z = 0$, we can expand $f, g, \Delta$ in the form
{\it e.g.} $f = f_{12} (w) + f_{11} (w)z + \cdots + f_{0}z^{12}$,
where $f_m$ is a polynomial of degree $m$ in $w$.  An explicit
analysis of $\gsu(n)$ models in this context was carried out in
\cite{mt-singularities}, and Weierstrass models for these theories
were found for $n \leq 20, n = 22,$ and $n = 24$, but no models were
identified for $n = 21, 23$.  Similarly, in \cite{KMT-I}, Weierstrass
models for elliptic fibrations over bases $B_2 = \F_1, \F_2$ with
Kodaira type $I_n$ singularities over the curves $S$ of
self-intersection $-1, -2$ in these bases were analyzed.  Anomaly
considerations suggest that in each case there are enough degrees of
freedom to tune an $I_{15}$ singularity, but solutions were only found
algebraically up to $n = 14$.

In general, such algebraically complicated problems arise whenever one
attempts to tune an $I_n$ or $I_n^*$ singularity.  For a complete
classification and enumeration of all elliptically fibered Calabi-Yau
threefolds with section, either a direct method is needed for
constructing a solution for the resulting set of polynomial equations
on the coefficients of $f, g$, or some more general theorem is needed
stating when this algebra problem has a solution.  This problem is
also in some cases apparently intertwined with the issue of
determining the discrete part of the gauge group, associated with
torsion in the Mordell-Weil group, as discussed in
\S\ref{sec:issues-abelian}.

\subsubsection{Codimension two singularities}
\label{sec:issues-matter}

The possible singularity types at codimension two are not completely
classified.  In most simple cases, a local rank one enhancement of the
gauge algebra gives matter that can be simply interpreted
\cite{Bershadsky-all, Katz-Vafa, Grassi-Morrison}.  For example, at a
point where an $I_n$ singularity locus crosses a $(0, 0, 1)$ component
of the discriminant locus $\Delta$ there is an enhancement to $I_{n +
  1}$ corresponding to matter in the fundamental representation of the
associated $\gsu (n)$.  In other cases, however, the singularities can
be more complicated.  Despite much recent progress in understanding
codimension two singularities and associated matter content
\cite{mt-singularities, Esole-Yau, Grassi-Morrison-2, Esole-Yau-II,
  Lawrie:2012gg, Grassi-hs-1, Hayashi:2014kca, Grassi-hs-2,
  Esole:2014bka}, there are still many aspects of codimension two
singularities, even for Calabi-Yau threefolds, that are still not well
understood or completely classified.  In principle, however, there
should be a systematic way of relating codimension two singularity
types to representation theory in the same way that the Kodaira-Tate
classification relates codimension one singularity types to Lie
algebras.

One particular class of codimension two singularities that is not as
yet systematically understood or classified are cases where the curve
$C$ that supports a Kodaira type singularity is itself singular.  For
simple singularity types, such as an intersection between two curves
--- which gives bifundamental matter --- or a simple intersection of
the curve with itself --- which for $\gsu (n)$ gives an adjoint
representation or a symmetric $+$ antisymmetric representation --- the
connection between representation theory and geometric singularities
is understood \cite{Sadov, mt-singularities}.  For more exotic
singularity types of $C$, however, there is as yet no full
understanding.  Analysis of anomalies in 6D theories \cite{kpt}
indicates that for any given representation ${\bf R}$, there is a
corresponding singularity that contributes to the arithmetic genus of
the curve $C$ through
\begin{equation}
g_{\bf R} = \frac{1}{12} 
(2 \lambda^2 C_{\bf R} + \lambda B_{{\bf R}} -\lambda A_{\bf R}) \,,
\label{eq:}
\end{equation}
where the anomaly coefficients $A_{\rm R}, B_{\rm R}, C_{\rm R}$ are
defined through (\ref{eq:bc-definition}).  For example, the ${\bf 20}$
``box'' representation of $SU(4)$ should correspond to a singularity
with arithmetic genus contribution 3 on the curve $C$; while a
potential realization of this representation through an embedding of
an $A_3$ singularity into a $D_6$ singularity was suggested at the
group theory level in \cite{mt-singularities}, the explicit geometry
of the associated singularity has not been worked out.  Without a
general theory for this kind of singularity structure, a complete
classification of EFS CY threefolds will not be possible.

\subsubsection{Mordell-Weil group and abelian gauge factors}
\label{sec:issues-abelian}

One of the trickiest issues that needs to be resolved for a complete
classification of EFS Calabi-Yau threefolds to be possible is the
problem of determining when additional nontrivial global sections of
an elliptic fibration over a given base $B_2$ can be constructed, and
explicitly constructing them when possible.  The construction of an
explicit Weierstrass model depends on the existence of a single global
section.  Using the fiber-wise addition operation on elliptic curves
(which corresponds to the usual addition law on $T^2$), the set of
global sections forms a abelian group known as the {\it Mordell-Weil}
group.  The Mordell-Weil group contains a free part $\Z^k$
of rank $k$, and can also have discrete torsion associated with
sections for which a finite multiple gives the identity (0 section).
The rank of the Mordell-Weil group determines the number of abelian
$U(1)$ factors in the corresponding 6D gauge group
\cite{Morrison-Vafa-II}.  In recent years there has been quite a bit
of progress in understanding the role of the Mordell-Weil group and
$U(1)$ factors in F-theory constructions and corresponding
supergravity theories \cite{Grimm-Weigand, Grimm-kpw, Park-Taylor,
  Park-abelian, Morrison-Park, Cvetic-gk, Mayrhofer:2012zy,
  Braun:2013yti, bmpw, Cvetic-Klevers, Braun:2013nqa, cgkp, bmpw-2,
  Cvetic-Klevers-2, Braun-fate, DPS, mt-sections}.  We review briefly
here some of the parts of this story relevant for constructing EFS CY
threefolds, and summarize some outstanding questions.

For a single $U(1)$ factor (rank 1 Mordell-Weil group), a general form
for the corresponding Weierstrass model was described by Morrison and
Park in \cite{Morrison-Park}.  It was shown in \cite{mt-sections} that
a Weierstrass model with a single section of this type can always be
tuned so that the global section, corresponding to a nontrivial
four-cycle in the total space of the Calabi-Yau threefold, becomes
``vertical'' and is associated with a codimension one Kodaira type
singularity giving a nonabelian gauge group factor in the 6D theory
with matter in the adjoint representation.  From the point of view of
this paper, this means that any model with a rank one Mordell-Weil group
can be constructed
by first tuning an $SU(2)$ or higher-rank nonabelian group on a curve
of nonzero genus, and then Higgsing the group using the adjoint matter
to give a residual $U(1)$ gauge group factor. This should in principle
make it possible to systematically construct all Calabi-Yau threefolds
with rank one Mordell-Weil group.

For higher rank, the story becomes more complicated.  Elliptic
fibrations with Mordell-Weil groups of rank two and three can be
realized by constructing threefolds where the fiber is realized in
different ways from the Weierstrass form \eq{eq:Weierstrass-1}
\cite{Klemm:1996ts, Aluffi-Esole}.  Explicit constructions of Weierstrass
models for general classes of threefolds with rank two and three
Mordell-Weil group were identified in \cite{bmpw, Cvetic-Klevers} and
\cite{Cvetic-Klevers-2} respectively, but there is no general
construction for models with Mordell-Weil rank higher than three.  CY
threefolds with much larger Mordell-Weil rank have been constructed;
it was shown in \cite{Martini-WT}, in particular, that for certain
$\C^*$ bases there is an automatic (``non-Higgsable'') Mordell-Weil
group of higher rank, with ranks up to $k = 8$.  It must be possible
to construct an elliptically fibered Calabi-Yau threefold over the
base $\P^2$ with Mordell-Weil rank seven; this follows from the
explicit construction in \cite{mt-singularities} of an $SU(8)$ model
with adjoint matter (with an $I_8$ singularity on a cubic curve),
which can be Higgsed to give $U(1)^7$ (though the explicit Higgsed
model has not been constructed).  It is also possible that an $SU(9)$
model with adjoint matter may exist on $\P^2$, which would give a
Mordell-Weil rank of 8.  It is not known whether all higher rank
Mordell-Weil groups can be constructed by Higgsing higher rank
nonabelian gauge groups; this would mean that the results of
\cite{mt-sections} or a single section could be generalized to an
arbitrary number of sections, so that all global sections could
simultaneously be tuned to vertical sections without changing $h^{1,
  1}$.  If this were true, it would lead to a systematic approach to
constructing all EFS CY threefolds with arbitrary Mordell-Weil rank,
but more work is needed to understand this structure for higher rank
models.  It is also known that the Mordell-Weil rank cannot be
arbitrarily high; for example, anomaly cancellation conditions in 6D
impose the constraint that the rank satisfies $k \leq 17$ when the
base is $\P^2$ \cite{Park-Taylor}, and this constraint can probably be
strengthened considerably.

Beyond the rank of the Mordell-Weil group, which affects the Hodge
numbers of the threefold formed by a particular Weierstrass model
through (\ref{eq:11}), the torsion part of the Mordell-Weil group is
also as yet incompletely understood.  For a complete classification of
EFS CY threefolds from Weierstrass models, a better understanding is
needed of what kinds of torsion in the Mordell-Weil group are possible
and how they can be tuned explicitly in Weierstrass models.  
In particular, while the Kodaira type dictates the Lie algebra of the
corresponding 6D theory, the gauge group $G$ may
take the form $\prod_{i}G_i/\Gamma$ where $G_i$ are the associated
simply connected groups and $\Gamma$ is a discrete subgroup dictated
by the torsion in the Mordell-Weil group.  We have not studied this
discrete structure in this work, but understanding it is necessary for
a full classification of EFS CY threefolds.
A
systematic discussion of Mordell-Weil torsion is given in
\cite{mx-torsion}.  Some examples of Mordell-Weil groups with torsion
are given, for example, in \cite{mt-sections}.

\subsection{Classifying all EFS Calabi-Yau fourfolds}
\label{sec:limitations-4}

The methods of this paper can be used to analyze elliptically fibered
Calabi-Yau manifolds of higher dimensionality, though there are
more serious technical and conceptual obstacles to a complete
classification of fourfolds or higher.  Elliptically fibered
Calabi-Yau fourfolds are of particular interest for F-theory
compactifications to the physically relevant case of four space-time
dimensions.  

The classification of minimal bases $B_2$ that support EFS Calabi-Yau
threefolds, which formed the starting point of the analysis here of
EFS CY3s with large $h^{2, 1}$, depended upon the mathematical
analysis of minimal surfaces and Grassi's result for minimal surfaces
that support an elliptically fibered CY threefold.  The analogous
results for fourfolds are less well understood.  In principle, the
mathematics of Mori theory \cite{Mori} can be used to determine the
minimal set of threefold bases that support EFS Calabi-Yau fourfolds,
but this story appears to be somewhat more complicated than the case
of complex base surfaces.  For fourfolds, the set of possible
transitions associated with tuning Weierstrass models include blowing
up curves as well as divisors, which further complicates the process
of analyzing the set of bases, even given the set of minimal bases.
Some basic aspects of these transitions are explored in
\cite{Klemm:1996ts, Grassi-network, Grimm-Taylor}.  At least in the
toric context, however, an analysis of CY fourfolds along the lines of
this paper seems tractable.  There has been some exploration of the
space of Calabi-Yau fourfolds with a toric description
\cite{Klemm:1996ts, Mohri, Kreuzer-Skarke-4, Knapp:2011wk,
  Bizet:2014uua}, and a complete enumeration of toric bases ${\cal
  B}_3$ with a $\P^1$ bundle structure that support elliptic fibrations
for F-theory models with smooth heterotic duals was carried out in
\cite{Anderson-WT}, along with a complete classification of non-toric
threefold bases with this structure.  A systematic analysis using
methods analogous to \cite{clusters, toric} of the space of all toric
bases that support elliptically fibered CY fourfolds seems tractable,
if computationally demanding.  Note that since over many bases there
are a vast number of different tunings, classifying the bases and
associated generic Weierstrass models is a much more tractable problem
than a complete classification of CY fourfold geometries.

\subsection{Further directions}
\label{sec:future}

The analysis initiated in this paper can in principle be continued to
substantially lower values of $h^{2, 1}$ before any of the issues
described in \S\ref{sec:limitations-3} become serious problems.  Even
outside the set of toric and $\C^*$ bases, the number of ways that the
Hirzebruch surfaces $\F_m$ with large $m$ can be blown up is fairly
restricted.  Algebraic problems with $I_n$ and $I_n^*$, nontrivial
Mordell-Weil groups, and exotic matter content are all issues that
become relevant only at lower values of $h^{2, 1}$.  Further work in
this direction is in progress, which may both reveal more about the
structure of elliptically fibered Calabi-Yau threefolds and may also
help provide specific situations in which the issues described in
\S\ref{sec:limitations-3} can be systematically addressed.  There are
a number of more general conceptual issues that can be addressed in
the context of this program, which we discuss briefly here.

\subsubsection{Hodge number structures}

The approach taken here, which in principle can systematically
identify all elliptically fibered Calabi-Yau threefolds that admit a
global section, is complementary to methods involving toric
constructions that have been used in many earlier studies of the
global space of CY threefolds.  The systematic analysis by Kreuzer and
Skarke \cite{Kreuzer-Skarke} of CY threefolds that can be realized as
hypersurfaces in toric varieties through the Batyrev construction
\cite{Batyrev} gives an enormous sample of Calabi-Yau threefolds whose
Hodge numbers have clear structure and boundaries.  The analysis of
elliptically fibered threefolds through Weierstrass models groups
the threefolds according to the base $B_2$ of the elliptic fibration,
and both simplifies the classification and enumeration of models and
enables the systematic study of non-toric elliptically fibered CY
threefolds.  The fact, observed in \cite{WT-Hodge, Martini-WT}, that
generic elliptic fibrations over both toric bases and a large class of
non-toric bases span a similar range of Hodge numbers, with similar
substructure and essentially the same boundary, suggests that these
sets of threefolds are not just a small random subset, but may in some
sense be a representative sample of all Calabi-Yau threefolds.  In
\cite{Candelas-cs}, Candelas, Constantin, and Skarke used the Batyrev
construction and the method of ``tops'' \cite{Candelas-font} to
analyze Calabi-Yau threefolds with an elliptic (K3) fibration
structure and identified certain patterns in the set of associated
Hodge numbers.  Some of these patterns are clearly related to the
transitions described through Weierstrass models as blow-ups and
tuning of gauge groups.  For example, the characteristic shift by
Hodge numbers of ($+ 1, -29$) is clearly seen from the Weierstrass
based analysis as the set of blow-up transitions between distinct
bases $B_2$.  Similarly, shifts such as $(+ 1, -17)$ can be seen as
arising from transitions on the full threefold geometry associated
with tuning an $\gsu (2)$ algebra on a $-1$ curve, \emph{etc}.  In
\cite{Candelas-cs}, another structure noted is the classification of
fibrations into ``$E_8$,'' ``$E_7$,'' \emph{etc.} types based on the way in
which the elliptic fiber degenerates along the base.  These correspond
precisely in the Weierstrass/base picture to the families of
threefolds that can be realized by blowing up points and tuning
additional gauge groups over the bases $\F_{12}, \F_8,$ etc..

One structure that is manifest in the Hodge numbers of Calabi-Yau
threefolds, however, which is not as transparent from the Weierstrass
point of view is the mirror symmetry of the set of threefolds, which
exchanges the Hodge numbers $h^{1, 1}$ and $h^{2, 1}$.  From the Batyrev point of
view, mirror symmetry has a simple interpretation in terms of the dual
polytope defining a toric variety used to construct a Calabi-Yau
manifold.  From the point of view of Weierstrass models of elliptic
fibrations over fixed bases, however, it seems harder to understand,
for example, how a blow-up transition with change in Hodge numbers $(+
1, -29)$ is related to a sequence of blow-ups that give a shift $(+
29, -1)$ and typically generate a full chain of divisors in the base
associated with a gauge group factor $E_8 \times F_4^2 \times (G_2
\times SU(2))^2$ \cite{toric, WT-Hodge}.  Understanding how these
two different approaches of toric constructions based on reflexive
polytopes and Weierstrass models on general bases can be brought
together to improve our understanding of mirror symmetry and the
structure of Hodge numbers for CY threefolds is an exciting direction
for further work.

\subsubsection{General
  Calabi-Yau's with large Hodge numbers}
\label{sec:role-EFS}

The results presented here add to a growing body of evidence that the
set of elliptically fibered CY threefolds with section may provide a
useful guide in studying general Calabi-Yau threefolds, and may in
fact dominate the set of possible Calabi-Yau threefolds.  While there
is as yet no clear argument that places any bound on the Hodge numbers
of a general Calabi-Yau threefold, several pieces of empirical
evidence seem to suggest that the CY threefolds with the largest Hodge
numbers may in fact be those that are elliptically fibered.  In this
paper we have shown that all Hodge numbers for known Calabi-Yau
manifolds that have $h^{2, 1}\geq 350$ are realized by elliptically
fibered threefolds.  
It seems natural to speculate that the threefolds constructed here may
constitute \emph{all} Calabi-Yau threefolds (elliptically fibered or
not)  that lie above this bound.
The results of \cite{WT-Hodge} suggest that more
generally, the outer boundary of the set of Hodge numbers for possible
CY threefolds may be realized in a systematic way by elliptically
fibered threefolds, and further empirical evidence from
\cite{Candelas-cs} also suggests that a large fraction of the models
in the Kreuzer Skarke database with large Hodge numbers are
elliptically fibered.  Since the methods of this paper do not depend
on toric geometry, it seems that this set of observations is not an
artifact of the toric approach, but rather that those threefolds
constructed using toric methods form a good sample, at least of those
threefolds that admit elliptic fibrations.  Other independent
approaches to constructing Calabi-Yau manifolds have recently given
further supporting evidence for the dominance of elliptically fibered
manifolds in the set of Calabi-Yaus.  In \cite{Gray-hl, Gray-hl-2},
the complete set of Calabi-Yau fourfolds constructed as complete
intersections in products of projective spaces were constructed.  From almost
one million distinct constructions it was found that 99.95\% admit
at least one elliptic fibration;
a similar analysis finds that 99.3\% of threefolds that are
complete intersections admit an elliptic fibration \cite{gray-private}.
Taken together, these results
suggest that it may be possible to prove that all Calabi-Yau
threefolds have Hodge numbers that satisfy the inequality $h^{1,
  1}+h^{2, 1} \leq 491 + 11 =502$.
Some initial exploration of one
approach to finding such a bound from the point of view of the
conformal field theory on the superstring world sheet has been
undertaken in \cite{Keller-Ooguri, Friedan-Keller}.

\subsubsection{New Calabi-Yau threefolds}
\label{sec:new}

In this paper we have identified three apparently new Calabi-Yau
threefolds, with Hodge numbers $(17, 371)$, $(18, 363)$, and $(19,
353)$.  We have performed a number of checks to confirm that these
models are consistent, which all work out,
so naively these appear to represent a new set
of Calabi-Yau threefolds.  Continuing the analysis of this paper to
lower Hodge numbers generates an increasingly large number of other
new threefolds, particularly as the bases involved themselves become
non-toric.  There are several questions that might be studied related
to the Hodge numbers found here that apparently describe new
Calabi-Yau threefolds.  One question is why these models do not appear
in the Kreuzer-Skarke database.  In particular, the new threefolds
identified here are elliptic fibrations over toric bases, so we might
expect in principle that they should appear in the Batyrev
construction.  One possible explanation may be that the structure of
the tuned $\gg_2$ gauge algebra that is common to all these
constructions somehow takes the full space outside the context of
hypersurfaces in toric varieties even though the base is still toric.

Another possible explanation for why they these new threefolds do not
appear in the Kreuzer-Skarke database, however, may be related to the
fact that they arise from tuning moduli in other Calabi-Yau threefolds
that have the same value of $h^{1, 1}$ (the threefolds with Hodge
numbers $(17, 377),$ $(18, 366)$, and $(19, 355)$ respectively),
associated with the enhancement of $\gsu (3)$ to $\gg_2$.  This means
that the geometric transitions associated with these tunings are less
dramatic than the other tunings and blow-ups since they do not
actually change the dimension of $H^{1, 1}$.  One possible scenario is
that these apparently new Calabi-Yau threefolds may actually represent
special loci in the moduli spaces of the corresponding $\gsu (3)$
structure threefolds, and might not actually represent topologically
distinct Calabi-Yau manifolds. This situation might be analogous to
the tuning of moduli in a base to give a $-2$ curve at a codimension
one space in the moduli space, which changes the structure of the Mori
cone but not the topology of the manifold.  Further study of the
detailed structure of these apparently new threefold constructions
goes beyond the methods developed in this paper but should in
principle be able to clarify this issue.

It would also be interesting to analyze the mirrors of these
apparently new threefolds.  A cursory check indicates that it is
difficult to construct the threefolds with the mirror Hodge numbers
where $h^{1, 1}$ and $h^{2, 1}$ are exchanged from tuned Weierstrass
models as elliptic fibrations.  This suggests either that the mirrors
may not be elliptically fibered or that the second explanation given
above is correct and that these are not actually topologically
distinct Calabi-Yau threefolds.  The former scenario would indicate
that the dominance of elliptic fibrations may be asymmetric in the
Hodge numbers.  Further understanding of these issues and construction
of additional new Calabi-Yau threefolds using these methods are an
interesting direction for further work.

\subsubsection{Uniqueness and equivalence of Calabi-Yau threefolds}
\label{sec:equivalence}

Another difficult problem on which the methods of this paper may be
able to shed some light is the question of when two Calabi-Yau
threefolds, given by different data, are identical.  In the
Kreuzer-Skarke database there are many examples of Hodge numbers for
which multiple toric constructions provide CY threefolds, as
illustrated in Table~\ref{t:Hodge-350}.  A priori, it is difficult to
tell when these threefolds represent the same complex manifold.
Wall's theorem \cite{Wall} states that when the Hodge numbers, triple
intersection numbers, and first Pontryagin class of the threefolds are
the same the spaces are the same as real manifolds, but even this does
not guarantee that two manifolds live in the same complex structure
moduli space.  The problem of telling whether two sets of triple
intersection numbers given in different bases are equivalent is also
by itself a difficult computational problem.  Thus, it is difficult to
tell whether two Calabi-Yau manifolds given, for example, by the toric
data in the Kreuzer-Skarke list, are identical.

The methods of this paper provide an approach that can resolve this
kind of question in some cases.  When the construction of an
elliptically fibered Calabi-Yau threefold over a given base with
specified Hodge numbers can be shown to be unique (up to moduli
deformation) using the Weierstrass methods implemented here, this
guarantees that any two CY threefolds that are both elliptically
fibered and share these Hodge numbers must be identical as complex
manifolds.  In particular, with the  exceptions of the Hodge number
pairs
$(10, 376)$ and $(19, 355),$
for all the Hodge numbers found in this paper with $h^{2, 1}> 350$,
there was a unique EFS CY threefold construction.  It follows that any
EFS CY threefolds with these Hodge numbers should be geometrically
identical as Calabi-Yau manifolds.  As an example, consider the
elliptically-fibered Calabi-Yau threefold with Hodge numbers (12,
462).  There are two distinct toric constructions of threefolds with
these Hodge numbers in the Kreuzer-Skarke database.  Both admit
elliptically fibrations.  As we have proved here in \S\ref{sec:12-1},
however, there is a unique construction of such a CY threefold, which
is realized by considering the generic elliptic fibration over a base
$\F_{12}^{[1]}$ given by blowing up the Hirzebruch surface $\F_{12}$
at any point not lying on the $-12$ curve $S$.  In principle,
continuing this kind of argument to lower Hodge numbers might be able
to significantly constrain the number of possible distinct Calabi-Yau
threefolds that can be realized using known constructions.  In
principle this line of reasoning can also be applied at a more refined
level by computing the triple intersection numbers for the manifolds
in question.  This approach may be able to distinguish some pairs of
elliptic fibration constructions with identical Hodge numbers, such as
the two constructions found here for threefolds with Hodge numbers
$(19, 353)$, or the generic elliptic fibrations over $\F_7$ and
$\F_8$, which both have Hodge numbers (10, 376).  Of course, however,
many CY threefolds are likely to admit multiple distinct elliptic
fibrations (as found for fourfolds in \cite{Gray-hl-2}), so in some
cases apparently distinct constructions of elliptic fibrations will
still give equivalent Calabi-Yau threefolds.  We leave further
exploration of these interesting questions to future work.
\vspace*{0.2in}

{\bf Acknowledgements}: We would like to thank Lara Anderson, David
Morrison for helpful discussions.  This research was supported by the
DOE under contract \#DE-FC02-94ER40818. In addition, SJ would like to acknowledge the generous support of the MIT Dean of Science Fellowship.

\end{document}